\documentclass[aps,prb,preprint,superscriptaddress]{revtex4-1}
\usepackage{graphicx}
\usepackage[T1]{fontenc}
\usepackage[utf8]{inputenc}
\usepackage{color}

\begin{document}

\title{Correlated electron-hole mechanism for molecular doping in organic semiconductors}

\author{Jing Li,$^1$ Gabriele D'Avino,$^{1,2}$ Anton Pershin,$^2$ Denis Jacquemin,$^{3,4}$ Ivan Duchemin,$^5$ David Beljonne,$^2$ Xavier Blase}
\affiliation{Grenoble Alpes University, CNRS, Inst. N\'{E}EL, F-38042 Grenoble, France \\
$^2$Laboratory for the Chemistry of Novel Materials, University of Mons, Place du Parc 20, BE-7000 Mons, Hainaut, Belgium \\
$^3$Laboratoire CEISAM - UMR CNR 6230, Université de Nantes, 2 Rue de la Houssini\`{e}re,
BP 92208, 44322 Nantes Cedex 3, France \\
$^4$Institut Universitaire de France, 1 rue Descartes, 75005 Paris Cedex 5, France \\
$^5$INAC, SP2M/L$\_$Sim, CEA/UJF Cedex 09, 38054 Grenoble, France }

\email{gabriele.davino@neel.cnrs.fr}
\email{xavier.blase@neel.cnrs.fr}

\date{\today}

\begin{abstract}
The electronic and optical properties of the paradigmatic F4TCNQ-doped pentacene in the low-doping limit are 
investigated by a combination of state-of-the-art many-body \emph{ab initio} methods accounting for environmental screening effects, and a carefully parametrized model Hamiltonian.
We demonstrate that while the acceptor level lies very deep in the gap, the inclusion of electron-hole interactions strongly stabilizes dopant-semiconductor charge transfer states and, together with spin statistics and structural relaxation effects, rationalize the possibility for 
room-temperature dopant ionization.
Our findings reconcile available experimental data, shedding light on the partial vs. full charge transfer scenario discussed in the literature, and question the relevance of the standard classification in  shallow or deep impurity levels prevailing for inorganic semiconductors.
\end{abstract}


\maketitle 


\section{Introduction}
\label{s:intro}

Doping of organic semiconductors (OSC) \cite{Wal07,Sal16} by introduction in the host matrix of strong electron- or hole-donating molecules has been shown to increase their electrical conductivity by orders of  magnitude, leading to  enhanced performances in organic light-emitting devices and photovoltaic cells.
However, in contrast to inorganic semiconductors where doping \cite{Sze} is understood to proceed via the formation of shallow dopant levels, 
the case of OSC remains controversial 
\cite{Sal12,Duo13,Pin13,Mit12,Men15,Png16,Kar16,Kang16}. 
Fundamental questions regarding the electronic structure of doped OSC and the evolution of the transport and optical properties with doping load are still open and it is presently  unclear  why very large dopant concentrations (a few percents)  are needed to boost their electrical conductivity.

\begin{figure}[ht]
\centering
\includegraphics[trim = 0 15 5 0, clip, width=0.7\textwidth]{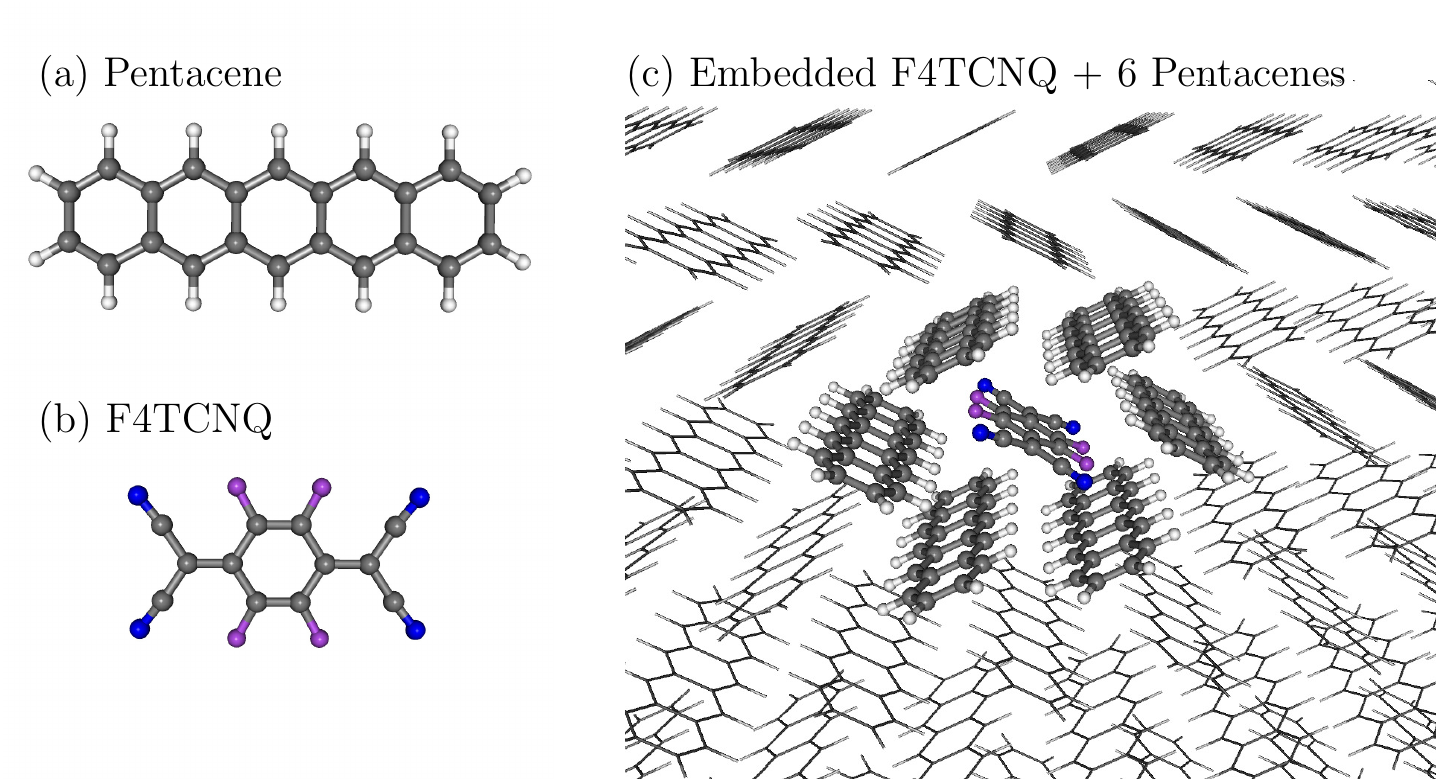}
\caption{(Color online)
Representation of the model system for molecular doping here studied with embedded many-body perturbation theory calculations.
Molecular structures of (a) pentacene and (b) F4TCNQ and of (c) the complex including the dopant surrounded by its six pentacene  neighbors (1+6 CPX, QM  region, ball-and-stick model). 
The 1+6 CPX is embedded in the pentacene crystal (MM region, wideframe 
representation).
}
\label{fig0}
\end{figure}

Experimental observations by Koch and coworkers point to  contrasting pictures \cite{Sal16,Men15}:
Full dopant-OSC charge transfer (CT) seems to be the paradigm for conjugated polymers \cite{Pin10,Gha15}, while spectroscopic evidences of partial CT, or dopant-OSC orbital hybridization, have been reported for different molecular systems \cite{Sal12,Men13}.
Focusing on the paradigmatic case of 2,3,5,6-tetrafluoro-7,7,8,8- tetracyanoquinodimethane (F4TCNQ) as a hole dopant in bulk pentacene (PEN, see Figure~\ref{fig0})  \cite{Ha09,Sal12,Kle12}, 
it was shown  that the introduction of the molecular dopant does not lead 
to any ultraviolet photoelectron spectroscopy (UPS) evidence
of singly-occupied levels in the pentacene gap, as expected according to the standard polaronic picture for polymers \cite{Bre85}.
Doping results instead in the emergence of two novel optical absorption lines at $\sim$1.2-1.4 eV, located below the 1.8 eV pentacene absorption onset \cite{Sal12}. 
These features have been ascribed to CT transitions pertaining to pentacene-F4TCNQ complexes on the basis of gas-phase density functional theory (DFT) calculations performed on a cofacial dimer  \cite{Sal12}. 
The emerging picture is that of strongly interacting molecular host-dopant pairs resulting in only \emph{partially ionized} dopants.
On the other hand, the scanning tunneling microscopy (STM) images by Ha and Kahn \cite{Ha09} instead show that isolated F4TCNQ in pentacene films are \emph{fully ionized} at room temperature,  an observation that has been rationalized on the basis of electrostatic modeling \cite{Top11}.

In this paper, we revisit the case of F4TCNQ-doped pentacene in the low-doping regime and analyze its electronic and optical properties with a combination of many-body \emph{ab initio} 
and model Hamiltonian electronic structure calculations, both explicitly accounting for electron-hole correlations.
We show that despite very deep acceptor levels in the pentacene gap, electron-hole interactions result in thermally accessible states with fully ionized F4TCNQ dopants. 
The broader picture obtained from our correlated electron-hole model describes doping as a competition between neutral and ionized dopants, passing through a narrow window of fractional CT. 
Our electronic structure calculations locate the pentacene-F4TCNQ system across this boundary, with structural molecular relaxation (polaronic effects) 
collapsing the system towards the full ionization of the dopant.
The optical absorption signatures of CT vs. ionized species in F4TCNQ-doped pentacene
and other doped OSC are analyzed in the light of recent experiments.

Our computational approach relies on state-of-the-art \emph{ab initio} electronic structure calculations using Green's function many-body perturbation theories within the  $GW$ \cite{Hed65,Oni02}
and Bethe-Salpeter equation (BSE) \cite{Han79,Oni02} formalisms. Extensive benchmarks against reference quantum-chemical calculations have demonstrated the accuracy of these approaches for calculating  quasiparticle energy levels \cite{Kni16,Kap16,Ran16b} and optical excitation energies \cite{Bru15,Jac15a}, 
BSE properly accounting for the long-range electron-hole interactions crucial for CT excitations \cite{Bau12,Duc12}.
The $GW$ formalism is embedded in a recently developed \cite{Li16} hybrid quantum/classical (QM/MM) scheme where many-body effects in the QM system are combined with an accurate discrete polarizable model accounting for the dielectric screening by the MM environment, known to largely affect the energies of charged and CT excitations \cite{Dav16rev}. 
This approach that proved to reproduce accurately the experimental photoemission gap and bulk (periodic) $GW$ calculations in pristine pentacene \cite{Li16} is here extended to optical excited states within the BSE framework. 

The paper is organized as follows. 
We first present in Section \ref{s:methods} the embedded \textit{ab initio} many-body formalism used, and the complementary model Hamiltonian that allows us to explore finite size effects and polaronic coupling. 
Our results are  presented in Section \ref{s:results}, 
followed by a discussion in Section \ref{s:disc}. 
After the conclusions and perspectives (Section \ref{s:conclu}), 
we gather in an Appendix convergence and validation tests performed on a 
small but representative subsystem.

\section{Methods}
\label{s:methods}

\subsection{Technical details of embedded many-body {\it ab initio} calculations}
\label{ss:details}

Our analysis relies on  first-principles many-body calculations within the 
framework of the hybrid QM/MM formalism presented in Ref.~\onlinecite{Li16}.
Our $GW$ and Bethe-Salpeter (BSE) calculations are performed with the Fiesta package \cite{Bla11a,Jac15a} that relies on Gaussian bases and Coulomb-fitting resolution-of-the-identity (RI-V) techniques. 
We adopt the triple-zeta plus polarization 6-311G(d) atomic basis set \cite{Pop80} combined with the Weigend Coulomb fitting \cite{Wei06} auxiliary basis set, ensuring convergence values of the photoemission gap and excitation energies  within a few  meV (see  Appendix \ref{s:app}). 
Input Kohn-Sham orbitals are generated by the NWChem 
package \cite{nwchem} modified so as to provide the DFT Kohn-Sham eigenstates and exchange-correlation matrix elements in the Gaussian atomic basis. 
Our $GW$ calculations are performed at the partially self-consistent (ev$GW$) level with update of the eigenvalues. Such an approach leads to improved accuracy for charged and neutral excitation energies \cite{Jac15a} 
and  cures the impact of the dramatic dependency of the starting Kohn-Sham HOMO-LUMO gap on the starting functional, as already documented for model donor-acceptor complexes. \cite{Sini11,Atalla13} For sake of consistency, we 
start with an hybrid functional PBEh($\alpha$=0.4),\cite{Per96}  
namely a functional with 40\% of exact exchange,
selected so as to match the starting Kohn-Sham gap with the $GW$ one (see below). BSE calculations are performed at the Tamm-Dancoff (TDA) level that  produces accurate energies for CT excitations.

The MM region is described by the charge response (CR) model by Tsiper and Soos 
\cite{Tsi01} in its \textsc{MESCAL} code implementation. \cite{Dav14}
This approach describes the anisotropic molecular response to electric fields  
in terms of induced atomic charges and induced dipoles, providing an 
accurate description of the static dielectric tensor of molecular solids. 
\cite{Tsi01_cpl,Dav16_jcp}
The molecular polarizability tensor is computed at DFT level [B3LYP functional, 6-311++G(d,p) basis set] and  atom-atom polarizabilities governing intramolecular charge flows are evaluated with semiempirical Hartree-Fock calculations (ZINDO parametrization).\cite{zindo}

Full technical details concerning the stability of our results with the starting DFT functional, convergency tests and comparison between full BSE and TDA calculations, are discussed in 
Appendix \ref{s:app}.

\subsection{Model Hamiltonian for molecular doping}
\label{s:model}

Our accurate many-body  \textit{ab initio} analysis is complemented with 
a generalized Mulliken model for intermolecular CT that allows 
us to describe systems of larger size and introduce structural relaxation (polaronic) 
effects.

For a single dopant (DOP) in the lattice of a host OSC we can represent the Hamiltonian on the basis of the neutral state $|\mathcal{N}\rangle$ and singlet-coupled full-CT states $|i_m\rangle$ with the electron populating the dopant LUMO and the hole in the HOMO$-m$ orbital ($m=0,1$) of the OSC site $i$. The electronic Hamiltonian reads:
\begin{eqnarray}
  H&=& \sum_{i,m}^{i\in \mathrm{OSC}} \varepsilon^\mathrm{CT}_{im} |i_m\rangle \langle i_m | + 
  \sum_{i,m}^{i\in \mathrm{OSC}} t_{im}^\mathrm{CT} \left( |\mathcal{N} \rangle \langle i_m | + |i_m \rangle \langle \mathcal{N} | \right ) \nonumber \\ 
&+&  \sum_{i,j}^{i,j\in \mathrm{OSC}} 
\sum_{m,n} t^\mathrm{h}_{im,jn} \left( |i_m \rangle \langle j_n | +  |j_n\rangle \langle i_m | \right ),
\label{e:H0}
\end{eqnarray}
where $\varepsilon^\mathrm{CT}_{im}$ are CT states energies, $t_{im}^\mathrm{CT}$ is the DOP-OSC charge transfer integral and $t^{h}_{ij}$ is the hole transfer integral between OSC sites.
Triplet excitations are described by the same Hamiltonian restricted to the subspace of CT states $|i_m\rangle$ with triplet spin pairing.
Hamiltonian \ref{e:H0} describes Coulombically-bonded and possibly delocalized electron-hole pairs and can be considered an extension of similar models successfully applied to the description of intra- \cite{terenziani06,terenziani08} and  intermolecular \cite{Dav16opv} CT.

Hamiltonian \ref{e:H0} applies to doping in molecular or polymer OSC and is here accurately parametrized from first principles for F4TCNQ-doped pentacene.
Diabatic CT states energies can be written as:  $\varepsilon^\mathrm{CT}_{im}=E^\mathrm{PEN}_{\mathrm{HOMO}-m}+E^\mathrm{DOP}_{\mathrm{LUMO}}+ P^\pm_{im}$.
\cite{Dav16rev}
Molecular orbital energies are calculated from gas-phase ev$GW$ calculations that yield
$E^\mathrm{PEN}_{\mathrm{HOMO}-1}=7.84$  eV,
$E^\mathrm{PEN}_{\mathrm{HOMO}}=6.36$  eV 
and $E^\mathrm{DOP}_{\mathrm{LUMO}}=3.76$ eV.
The polarization energy $P^\pm_{im}$ accounts for electrostatic and screening effects in the solid-state.
$P^\pm_{i0}$ is evaluated for each CT state with CR calculations, whose results are extrapolated in the infinite crystal limit. 
Polarization energies of CT states with the hole in the pentacene HOMO$-1$ are set to $P_{i1}^\pm=P_{i0}^\pm-0.3$ eV.

Hole transfer couplings $t^{\mathrm{h}}$ are computed at the DFT level (PBE0 functional, 6-31G(d) basis) with the  projective method.\cite{valeev06}
The same approach proved to be strongly dependent on the functional for pentacene-F4TCNQ CT couplings $t^\mathrm{CT}$, for which we instead applied a multi-state generalized Mulliken-Hush diabatization \cite{cave96} scheme based on a post-Hartree-Fock description [SCS-CC2/def-SV(P)] of the 
ground and excited states \cite{hellweg08}.
The sign of CT couplings is usually not attainable with standard
quantum chemistry approaches because of the arbitrariness of the phase of $\pi$ molecular orbitals.
We overcome this limitation by using a fictitious $s$ orbital placed above the molecular plane to probe the phase of frontier orbitals and then impose consistent phase relationships among all the molecules in the sample.  

\begin{figure}[h!]
\includegraphics[width=0.45\columnwidth]{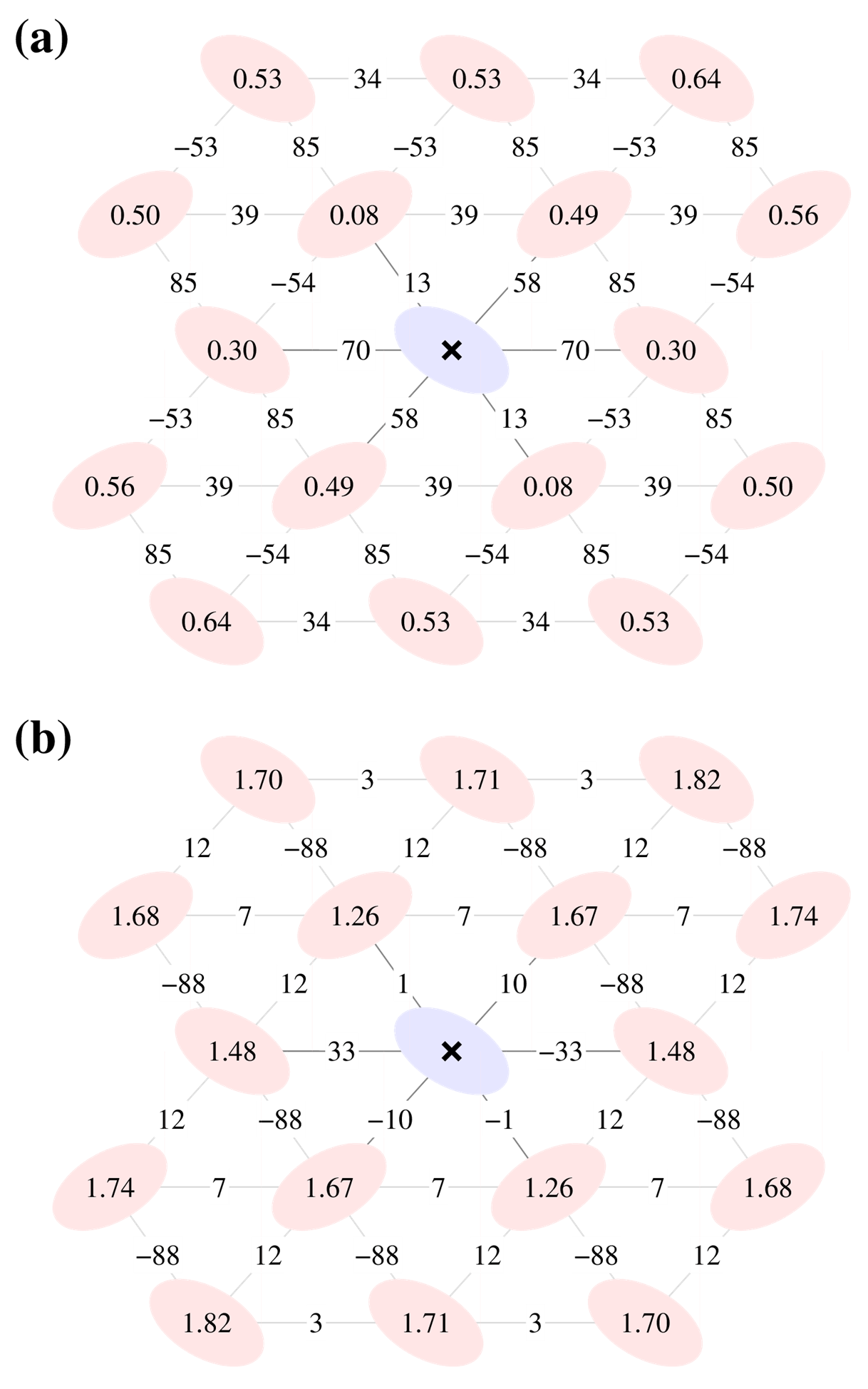} 
\caption{
(a) Energies of diabatic CT states ($\varepsilon^{CT}_{i0}$ in eV) with electrons localized 
on the central F4TCNQ (cross) and hole in the HOMO of each pentacene molecule $i$. 
Intermolecular couplings ($t^{CT}_{i0}$ and $t^h_{i0,j0}$ in meV) are annotated 
on the respective bonds. 
(b) Same parameters  as in the previous panel but for CT states with the hole in the pentacene HOMO$-1$, 
$\varepsilon^{CT}_{i1}$,  $t^{CT}_{i1}$ and $t^h_{i1,j1}$.
}
\label{CTene}
\end{figure}

\begin{figure}[h!]
\centerline{\includegraphics[width=0.4\columnwidth]{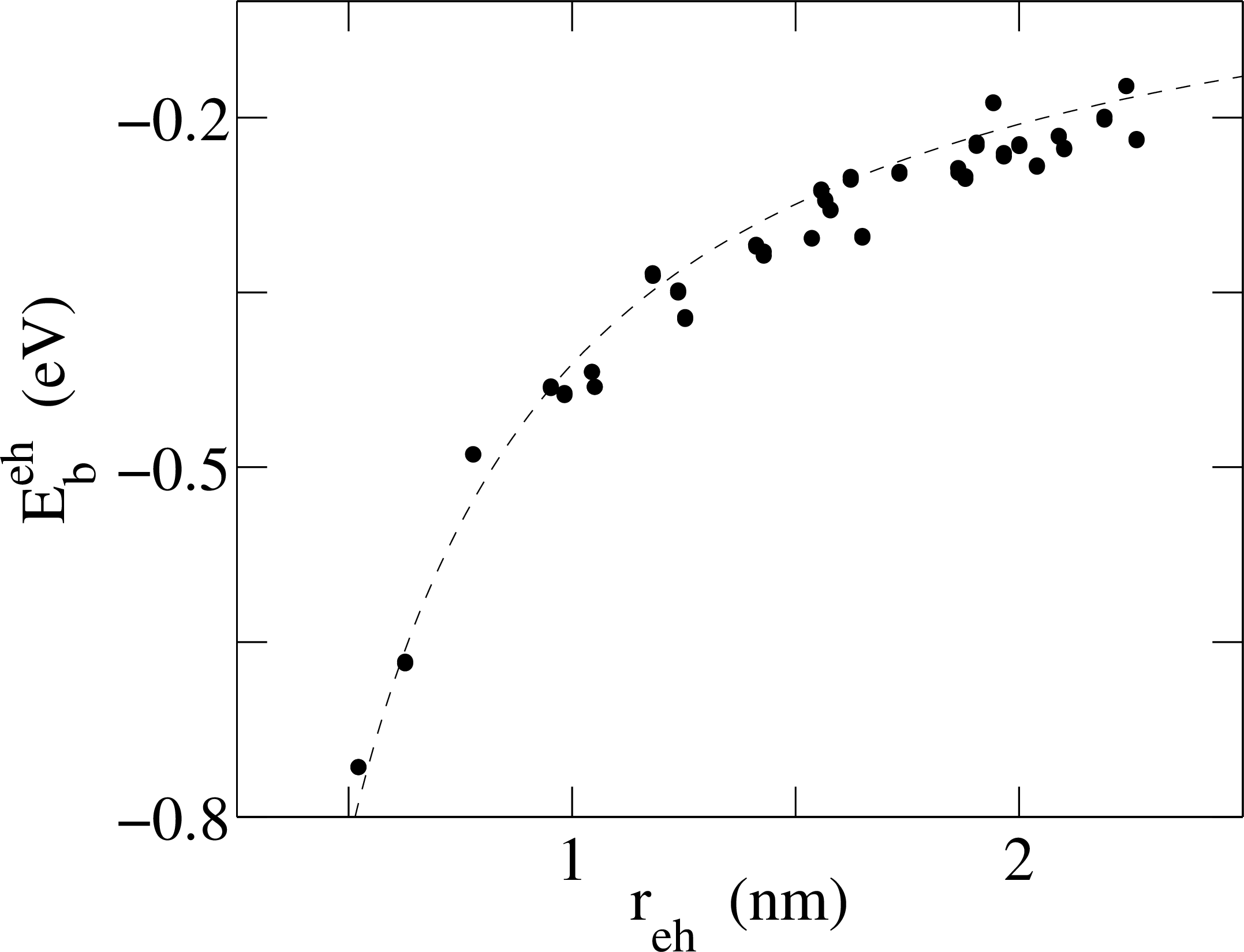}}
\caption{Electron-hole distance dependence of the exciton binding energy $E^\pm_b$ of
localized (diabatic) CT states $E^\pm_b=P^\pm - P^+ -P^- $.
$P^+$, $P^- $ and $P^\pm$ are the polarization energies for holes, electrons, and ion pairs computed with the CR model, all extrapolated in the infinite bulk limit. 
The exciton binding energy approximately follows a screened Coulomb potential
$-(\varepsilon_r r_{eh})^{-1}$ (dashed line for $\varepsilon_r=3.5$). }
\label{xcb}
\end{figure}

Figure \ref{CTene} provides a graphical summary of the parameters entering Hamiltonian 
\ref{e:H0}, i.e. the CT couplings and the energies of localized (diabatic) CT states annotated on the lattice of doped pentacene. 
Figure \ref{xcb} shows the electron-hole distance dependence of the exciton binding energy of diabatic CT states from CR calculations, which closely follows a screened Coulomb potential even at relatively short distance.

Hamiltonian \ref{e:H0} can be extended to account for intramolecular structural 
relaxation upon charging within the framework of the Mulliken-Holstein model. 
We hence introduce one effective mode per molecule with coordinate $q_i$, here treated 
within the adiabatic (Born-Oppenheimer) approximation, 
linearly modulating the energies of frontier molecular orbitals.
The electronic Hamiltonian in the presence of Holstein coupling is formally equivalent to Equation~\ref{e:H0} with  $E^\mathrm{DOP}_{\mathrm{LUMO}}\rightarrow E^\mathrm{DOP}_{\mathrm{LUMO}} + q_1$ ($i=1$ labels the DOP site)
and $E^\mathrm{PEN}_{\mathrm{HOMO}-m} \rightarrow E^\mathrm{PEN}_{\mathrm{HOMO}-m} + q_i$, with $i=2,\dots N$, being $N$ is the number of molecular sites.
The total energy includes the harmonic elastic contribution from molecular deformation,
\begin{equation}
E_{harm}=\frac{1}{4\lambda^-} q_1^2 + \frac{1}{4\lambda^+} \sum_i^{i \in OSC} q_i^2,
\end{equation}
where $\lambda^+$ and $\lambda^-$ are the polaron binding energies for hole and electrons
on OSC and DOP sites, respectively. 
These quantities have been calculated for PEN ($\lambda^+=52$ meV) and 
F4TCNQ ($\lambda^-=140$ meV) at the DFT level (PBE0 functional, 6-311G(d) basis)
using differences of total energy obtained at the molecular geometries fully relaxed in the neutral and charged state ($\Delta$SCF scheme).

\section{Results}
\label{s:results}

\subsection{Embedded $GW$ and BSE calculations of F4TCNQ-doped pentacene}
\label{s:res1}

The model system investigated here considers  F4TCNQ substitutional defects in the pentacene crystal lattice \cite{Ha09} -- see Figure \ref{fig0}. Within our hybrid formalism, we describe a supramolecular complex (CPX) formed by one F4TCNQ molecule surrounded by its first shell of six pentacene neighbors (1+6 CPX henceforth) at the $GW$/BSE level. 
This CPX is then embedded into the pentacene crystal described within the charge response (CR) model \cite{Tsi01,Dav14}, which provides an accurate description of the anisotropic static dielectric response of molecular crystals \cite{Dav14}.
F4TCNQ adopts the same position and orientation as that of the replaced pentacene molecule and its geometry has been optimized in vacuum at the CCSD level. The pentacene structure is taken from X-ray diffraction data for the vapor-grown polymorph \cite{Sie01}. 
We stress that such an approach goes significantly beyond previous DFT electronic structure calculations \cite{Sal12}, in terms of method accuracy, QM system size and account of an atomistic polarizable embedding.

\begin{figure}[ht]
\centering
\includegraphics[trim = 10 10 10 10, clip, width=0.7\textwidth]{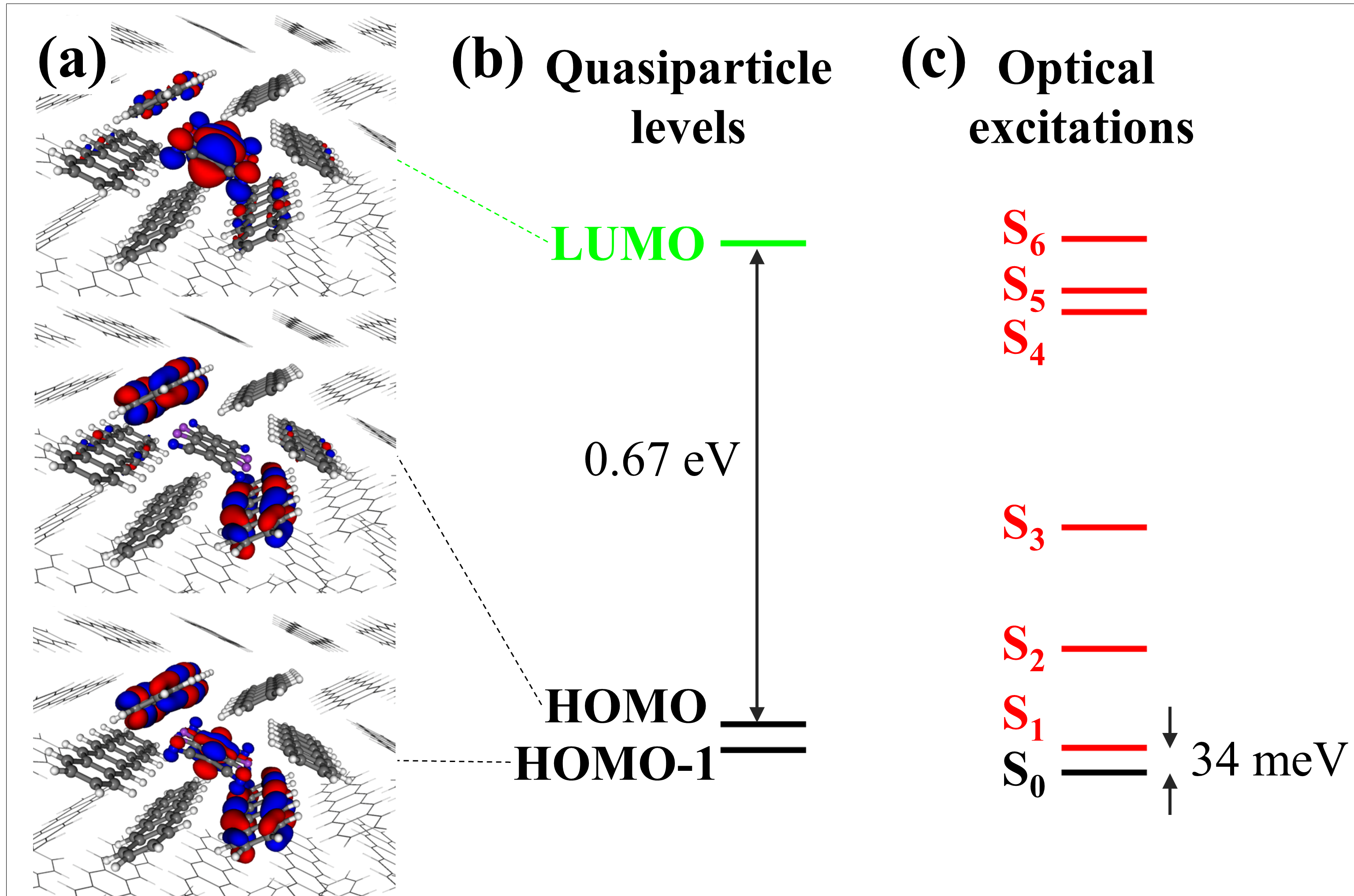}
\caption{(Color online)
(a) Isocontour representation of the frontier molecular orbitals corresponding to the (b) $GW$ energy levels computed for the embedded 1+6 CPX.
(c) BSE optical absorption spectrum of the embedded CPX. 
The striking difference between the HOMO-LUMO gap and the energy of the corresponding $S_1$ transition quantifies the strength of the electron-hole binding energy. }
\label{fig1}
\end{figure}

The first outcome of the present calculations, namely the accurate determination of the 1+6 CPX HOMO-LUMO gap in a proper dielectric environment, illustrates the striking difference between the physics of OSC doping and that prevailing in inorganic semiconductors where shallow impurity levels are located within a few dozen of meVs from the band edges. 
As shown in Figure \ref{fig1}(b), the $GW$ HOMO-LUMO gap of the CPX is indeed found to be 0.67 eV, dramatically larger than room temperature thermal energy, clearly evidencing that the standard theory accepted for inorganic semiconductors does not apply to OSC.

The analysis of the frontier orbital isocontours in Figure~\ref{fig1}(a) reveals that while the 1+6 CPX LUMO can be approximately identified as the LUMO of the F4TCNQ molecule, the two highest occupied orbitals have instead a supramolecular character, mostly involving the HOMOs of a specific pair of equivalent pentacene molecules.
To better understand the origin of the HOMO-LUMO energy gap, it is useful to re-evaluate it in the absence of hybridization, namely by successively considering either a single F4TCNQ or a single pentacene molecule in the QM region. 
The resulting $GW$ gap between pentacene HOMO and F4TCNQ LUMO in absence of dopant-host interactions, but with the proper dielectric environment, is found to be 0.45 eV.
Hybridization therefore contributes to the HOMO-LUMO gap of the CPX, yet the latter is mostly sourced by the energy mismatch between the pentacene ionization potential (IP) and the F4TCNQ electron affinity (EA).
The partial hybridization within the 1+6 CPX manifests in the frontier molecular orbitals shown in Figure \ref{fig1}(a) and in an appreciable CT 
in the DFT ground state, with a net charge on F4TCNQ $Q_\mathrm{DOP}=-0.25e$ as estimated from electrostatic potential fitting (ESP) atomic charges.

\begin{figure}[ht]
\centering
\includegraphics[trim = 10 10 10 10, clip, width=0.6\textwidth]{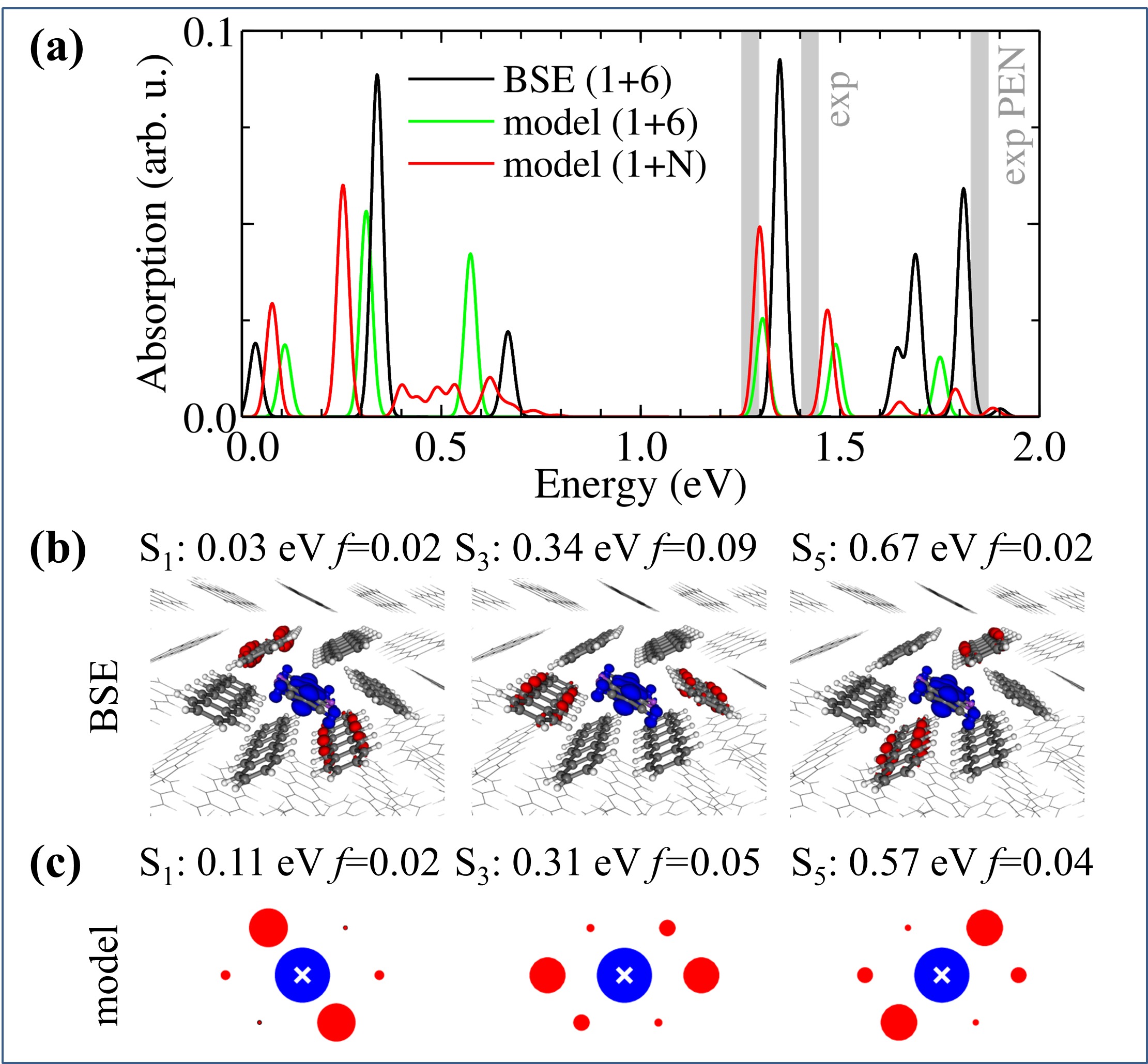}
\caption{(Color online) 
(a) Optical absorption spectrum of F4TCNQ-doped pentacene computed 
\emph{ab initio} (BSE) and with the model Hamiltonian  \ref{e:H0} 
for the 1+6 and 1+$N$ complexes at the geometry of the neutral molecules. 
Gray bars mark bands observed in experiments \cite{Sal12}.   
Hole (red) and electron (blue) density for the lowest-energy 
optically allowed excitations from (b) BSE and (c) model calculations. 
The BSE representation corresponds to the hole-averaged electron-density and electron-average hole-density associated with
the electron-hole two-body $\psi({\bf r}_e,{\bf r}_h)$ BSE eigenstates.
Transition energies and oscillator strengths $f$ are annotated.}
\label{fig2}
\end{figure}

We now turn to  optical excitations as obtained within the BSE formalism.
The (screened) electron-hole interaction  dramatically lowers the optical gap as compared to the HOMO-LUMO gap, as evidenced by comparing Figure~\ref{fig1}(b) and \ref{fig1}(c). 
The resulting CPX optical absorption spectrum is shown in Figure~\ref{fig2}(a). 
The most salient feature is that the lowest singlet excitation energy (S$_1$) is found to be extremely low in energy, namely 34 meV above the ground state. 
The analysis of the corresponding BSE electron-hole two-body eigenstate, represented in Figure~\ref{fig2}(b), reveals that S$_1$ is an excitation of CT character taking place between HOMO and LUMO of the CPX. 
Additional absorption peaks in the 0.3-0.6 eV range (S$_3$, S$_5$) are also associated with transitions mostly from the HOMO levels of other pairs of equivalent pentacene molecules in the CPX.
Such peaks fall in the mid-infrared spectral region that has so far not been investigated experimentally.\cite{Sal12}
The spectrometer employed in Ref.~\onlinecite{Sal12} was unable to measure absorption below 0.5 eV.\cite{ingo}

The  BSE optical spectrum also closely reproduces the fundamental absorption of pristine pentacene at $1.85$ eV and, most interestingly, displays further new features in the 1.3--1.6 eV energy range that may correspond to the sub-bandgap peaks observed experimentally \cite{Sal12}.
The analysis of the contributing levels reveals that these excitations correspond mainly 
to CT transitions from the manifold of pentacene HOMO$-1$ orbitals to the F4TCNQ LUMO.
We will come back to this point in Section~\ref{s:disc}.

\subsection{Model validation and size effects} 
\label{s:res2}

Even though  based on an accurate \textit{ab initio} many-body framework which is at the forefront of what can be achieved today in terms of system size and complexity, the present $GW$/BSE calculations on the 1+6 CPX miss to incorporate potentially important effects associated with the delocalization of the transferred hole over pentacene molecules beyond the first shell of neighbors and local structural relaxation or polaronic effects.
To extend the reach of our analysis to larger system sizes, we resort to the generalized Mulliken model for intermolecular CT presented in Section \ref{s:model}.

The  diagonalization of Hamiltonian \ref{e:H0} with parameters specific to F4TCNQ-doped pentacene, as described in Section \ref{s:model}, yields ground and excited states for systems large enough to converge the quantities of interest with respect to the number of pentacene molecules (1+$N$ CPX  henceforth).
Our model calculations yield a ground state of moderate CT,  consistently with the \textit{ab initio} findings.
The charge on the dopant  mildly increases from $-0.15e$ in the 1+6 CPX to $-0.24e$ 
for the 1+$N$ CPX.
This trend reflects the reduction of the gap due to hole delocalization in pentacene.

The absorption spectrum computed with Hamiltonian~\ref{e:H0} for the 1+6 CPX is compared to BSE results in Figure~\ref{fig2}(a).
Again, the low-energy region is characterized by three CT transitions to 
electronic excited states where the hole is equally shared on the HOMOs of pairs of symmetry-equivalent pentacene molecules.
In the 1.2--1.8 eV energy span, we predict three other absorption peaks corresponding to CT excitations where the hole lies in the pentacene HOMO$-1$ orbitals. 
The agreement with BSE is excellent for both 
the relative energy and intensity of the electronic transitions and the shape of the excited-state electron-hole maps, see Figure~\ref{fig2}(b-c). This provides an important validation step for our model.

Finite-size effects are also addressed in Figure~\ref{fig2}(a), where we compare the
absorption spectra of the 1+6 and 1+$N$ CPX calculated with Hamiltonian~\ref{e:H0}.
Upon extending the size of the system (1+$N$ CPX), we observe 
a small shift to lower energies of the lowest two energy bands
due to the delocalization-induced stabilization of S$_1$ and S$_3$ and many additional transitions in the 0.4-0.8 eV range, signaling the presence of a dense manifold of delocalized CT states. 
Size effects affect in a similar way CT states arising from pentacene HOMO$-1$ in the 1.3-1.9 eV range.

On a more general vein, Hamiltonian \ref{e:H0} describes a smooth transition from a neutral ($Q_\mathrm{DOP}=0$) to a fully ionized dopant ($Q_\mathrm{DOP}=-e$) as a function of the difference between the IP and EA of the two components (with all other parameters fixed), Figure \ref{fig3}(a).
The width of the intermediate-ionicity window primarily depends on the magnitude of 
the couplings between the pentacene HOMO and the F4TNCQ LUMO ($t_{i0}^\mathrm{CT}$),
with large values favoring a mixed-valence ground state and zero or full CT in the opposite case.
We also remark that the ground-state value of $Q_\mathrm{DOP}$ (0 K) can substantially differ from its thermal average at 300 K in the crossover regime, mostly owing to triplet excitations which are full-CT states of threefold multiplicity.

\subsection{Structural relaxation and polaronic effects}
\label{s:res3}

The picture emerging so far depicts F4TCNQ-doped pentacene as a system close to the neutral-ionized boundary, as testified by the presence of CT excitations below 0.1 eV.
Such a small gap is commensurate with the polaron binding energies for F4TCNQ-pentacene,
$\lambda^+ + \lambda^- =0.2$ eV (see Section \ref{s:model}), 
calling for the inclusion of vibrational effects in the analysis.

\begin{figure}[ht]
\centering
\includegraphics[trim = 20 10 10 10, clip, width=0.6\textwidth]{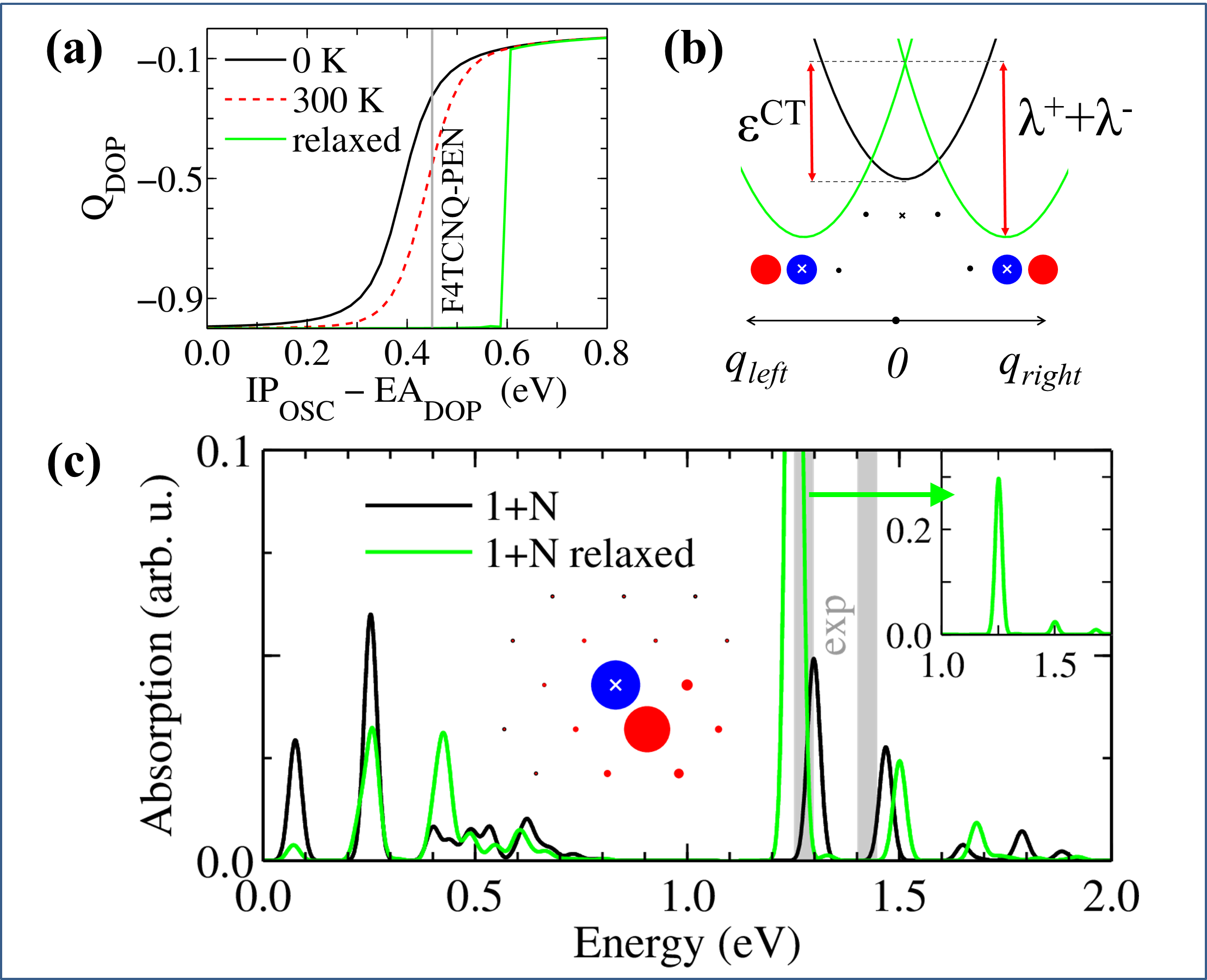}
\caption{(Color online)
(a) Dopant charge  in the ground-state (0 K) and at room-temperature (300 K) 
as a function of the relative strength of  the dopant-semiconductor pair. 
The temperature-independent green curve refers to a system at relaxed geometry.
(b) Sketch of the diabatic potential energy surfaces  illustrating the Jahn-Teller 
instability for an hypothetical 1+2 system.
(c) Optical absorption of a (un)relaxed 1+$N$ system. 
The inset shows the charge distribution of the vibrationally-relaxed ground state with 
a fully ionized dopant ($Q_\mathrm{DOP}=0.99e$).
All results from Hamiltonian \ref{e:H0}.
}
\label{fig3}
\end{figure}

We hence extended the model to account for one effective Holstein mode per molecule linearly coupled to the site charge as  described  in  Section \ref{s:model}.
The optimization of the ground state energy with respect to the set of intra-molecular coordinates leads to a qualitatively different \emph{symmetry-broken ground state of full-CT character}, shown in the inset of Figure \ref{fig3}(c).
Indeed, because of the two low-lying degenerate CT states at the undistorted molecular geometries, a structural (Jahn-Teller like) instability develops in the system, leading to the collapse of the hole on one of the nearest pentacene neighbors  -- see Figure \ref{fig3}(b). 
Structural relaxation therefore favors full dopant ionization and suppresses states of 
fractional CT. This is shown by the $Q_\mathrm{DOP}$ vs. $(\mathrm{IP_{OSC}-EA_{DOP}})$ curve in Figure \ref{fig3}(a) computed at relaxed molecular geometries (green line),
presenting an abrupt step-like transition whose turning point is shifted to higher $(\mathrm{IP_{OSC}-EA_{DOP}})$ with respect to the rigid-molecule case. 
Hole localization is also observed upon introducing Gaussian energetic disorder in the model, which lifts the equivalence of pentacene sites. 
Structural relaxation and disorder are expected to concur in localizing charges in real (experimental) systems.

The optical absorption of the 1+$N$ CPX at the relaxed and unrelaxed geometry, both shown in Figure \ref{fig3}(c), are only apparently similar.
For instance, the lowest energy transition at 76 meV now leads the relaxed ground state  to another CT state with the hole localized on the HOMO of the pentacene located on the opposite side of the dopant.
Though located at similar energy, it has a completely different nature compared to 
the optical transition S$_1$ in Figure~\ref{fig2}(c),
which was a CT transition transferring one electron from pentacene to F4TCNQ
from an almost neutral ground state.
The most relevant feature of the absorption of the vibrationally-relaxed CPX is, 
however, the very intense band at 1.25 eV, which falls very close in energy 
to the full-CT (pentacene HOMO$-1$ $\rightarrow$  F4TCNQ LUMO) transition discussed above for the unrelaxed system. 
This quasi-coincidence is a specific feature of the pentacene-F4TCNQ CPX and of its nearly vanishing optical gap.
Indeed, this band corresponds to a transition from the CT ground state in the Figure \ref{fig3}(c) (inset), with the hole in the pentacene HOMO, to another full-CT excited state where the hole is in the HOMO$-1$ of the same molecule.
This band owns its large absorption intensity to the off-diagonal element of the dipole moment operator associated with such an intramolecular reshuffling of the hole density \cite{cave96}.
In the most common nomenclature, the  band at 1.25 eV corresponds to a  \emph{polaronic transition} of the charged pentacene in the presence of a counter-electron on F4TCNQ.

\clearpage
\section{Discussion}
\label{s:disc}
The experimental observation of sub-bandgad absorption bands in F4TCNQ-doped pentacene 
can be rationalized both in terms of CT bands in weakly interacting complexes, or 
as polaronic features of ionized pentacene.
Our accurate analysis leads us to assign these sub-bandgap features to the pentacene radical cation.
This assignment is further corroborated by spectroscopic measurements on solutions of 
TIPS-pentacene oxidized by the very strong electron acceptor  FeCl$_3$ \cite{Sak10}.
The strikingly coincident double peak observed at 1.2--1.4 eV 
both in F4TCNQ- \cite{Sal12} and FeCl$_3$-doped (TIPS-)pentacene \cite{Sak10}
is a very strong indication of the same origin of these features, namely two vibronic replica 
of the same polaronic transition \cite{Sak10}.
The experimental band at $\sim$1.4 eV may also have a contribution from a characteristic absorption 
of the F4TCNQ anion \cite{Men15}.

Extensive spectroscopic (photoemission, optical and vibrational)  investigations by Koch
and co-workers allowed identifying two different scenarios for doped molecular crystals,
characterized by strong orbital hybridization and partial CT, and conjugated polymers,
mostly exhibiting full ionization of dopant impurities.\cite{Sal12,Men13,Men15,Sal16}
Our accurate theoretical analysis, tightly connected to experimental evidence,
concludes that F4TCNQ-doped pentacene represents an exception to this empirical rule, at least in the low-doping regime targeted by our calculations.
We further note that at least one case of partial CT has also been reported in conjugated  polymers.\cite{Gha15}

We emphasize that the absence of intragap features in the photoemission spectra of doped pentacene is not inconsistent with full dopant ionization and with the presence of pentacene cation sub-bandgap optical excitations.
Indeed, the intramolecular charging energy, associated with photoemission, is expected to split the singly and doubly occupied levels of pentacene, bringing the UPS features of the cation (polaron) deep in the valence band, as very recently shown for C$_{60}$ fullerenes \cite{Win15} and a narrow-band polymer  \cite{Png16}.
This charging energy is instead irrelevant for optical transitions, namely excitations localized on the pentacene cation that do not affect the molecular charge.

Vibrational and optical data on  other molecular systems, 
such as quaterthiophene (T4) \cite{Men15} or  2,7-didecyl[1]benzothieno-[3,2-b][1]benzothiophene (BTBT) \cite{Men13}, 
instead provide clear evidence in support of the weak CT CPX scenario.
This is rationalized by the curve in Figure \ref{fig3}(a), when we consider the
$\sim$0.4 eV larger IP of T4 \cite{Men15} and BTBT \cite{Tsu16} with respect to pentacene 
($\mathrm{IP}_{PEN}=4.9$ eV) \cite{Sal12}.
Further confirmations of weak dopant ionization come also from very recent correlated-electron calculations targeting the 1:1 F4TCNQ:BTBT mixed-stack crystal.\cite{Delc17}
A less efficient charge injection in the OSC matrix does also rationalize 
the 2-3 orders of magnitude lower conductivity in doped T4 \cite{Men15} 
and BTBT \cite{Men13} with respect to doped pentacene \cite{Kle12},
despite the comparable or higher charge carrier mobility in the formers.

Doped polymers can also be described by  Hamiltonian \ref{e:H0}, although  quantitative differences in its parameters should be expected with respect to molecular systems.
On the basis of approximate empirical parametrization, we suggest that the full dopant ionization observed experimentally in the large majority of doped polymers should be ascribed to their generally lower IP resulting from large bandwidths  
(e.g. in poly(3-hexylthiophene) IP=4.6 eV \cite{Men15}, bandwidth $\sim3.5$ eV \cite{Col11}).
Bandwidths larger than the charging energy should also grant validity of the single particle picture in most conjugated polymers, hence explaining the observation of intra-gap states in UPS arising from polaronic relaxation \cite{Bre85}.

\section{Conclusion and perspectives}
\label{s:conclu}
In summary, we have developed a multifaceted approach combining state-of-the-art electronic structure methods and applied this formalism to gain a comprehensive picture of the electronic and optical properties of organic semiconductors in the low doping regime.
Our analysis is based on novel many-body {\it ab initio} techniques that allow the accurate determination of charged and optical excitations in heterogeneous molecular systems, fully accounting for the complex non-local physics of interacting electrons and for the dielectric screening provided by the molecular environment.
First principles calculations are also employed to parametrize a general model Hamiltonian  for doped organic semiconductors covering the entire spectrum of possibilities between neutral and fully ionized dopants.

By explicitly considering the prototypical case of F4TNQ-doped pentacene, our calculations show that even for acceptor dopant levels lying fairly deep into the gap,
full ionization may still be possible thanks to the concurrent effects of electron-hole interaction, spin statistics and polaronic relaxation, yet leading to charge carriers that are strongly bound to the parent dopants.
Our analysis confirms the common belief that the difference between the IP of the semiconductor and the dopant EA is an important parameter but certainly not the only one. 
Indeed, other quantities such as electron-hole interaction, dopant-host CT integrals and the coupling to vibrations are found to be key for dopant ionization.
All these quantities can be strongly dependent on the material and on the morphology, calling for a detailed analysis of structure-property relationships.
The proposed approach provides a robust framework to such a scope.

On more general grounds, this study puts the accent on the central role of electron-hole interaction and polaronic effects in favoring dopant ionization, two effects 
that are missed in the well-established theories for electrical doping in inorganic semiconductors, whose applicability to the case of organic systems is here called into question.
In contrast to the picture prevailing for doped inorganic semiconductors, the explanation for the high conductivity in heavily doped organics should thus be sought beyond independent-electron theories.

\appendix

\section{Test calculations with the embedded many-body approach}
\label{s:app}

\subsection{Stability with respect to starting Kohn-Sham eigenstates and basis set}

\begin{figure}
\includegraphics[width=0.4\columnwidth]{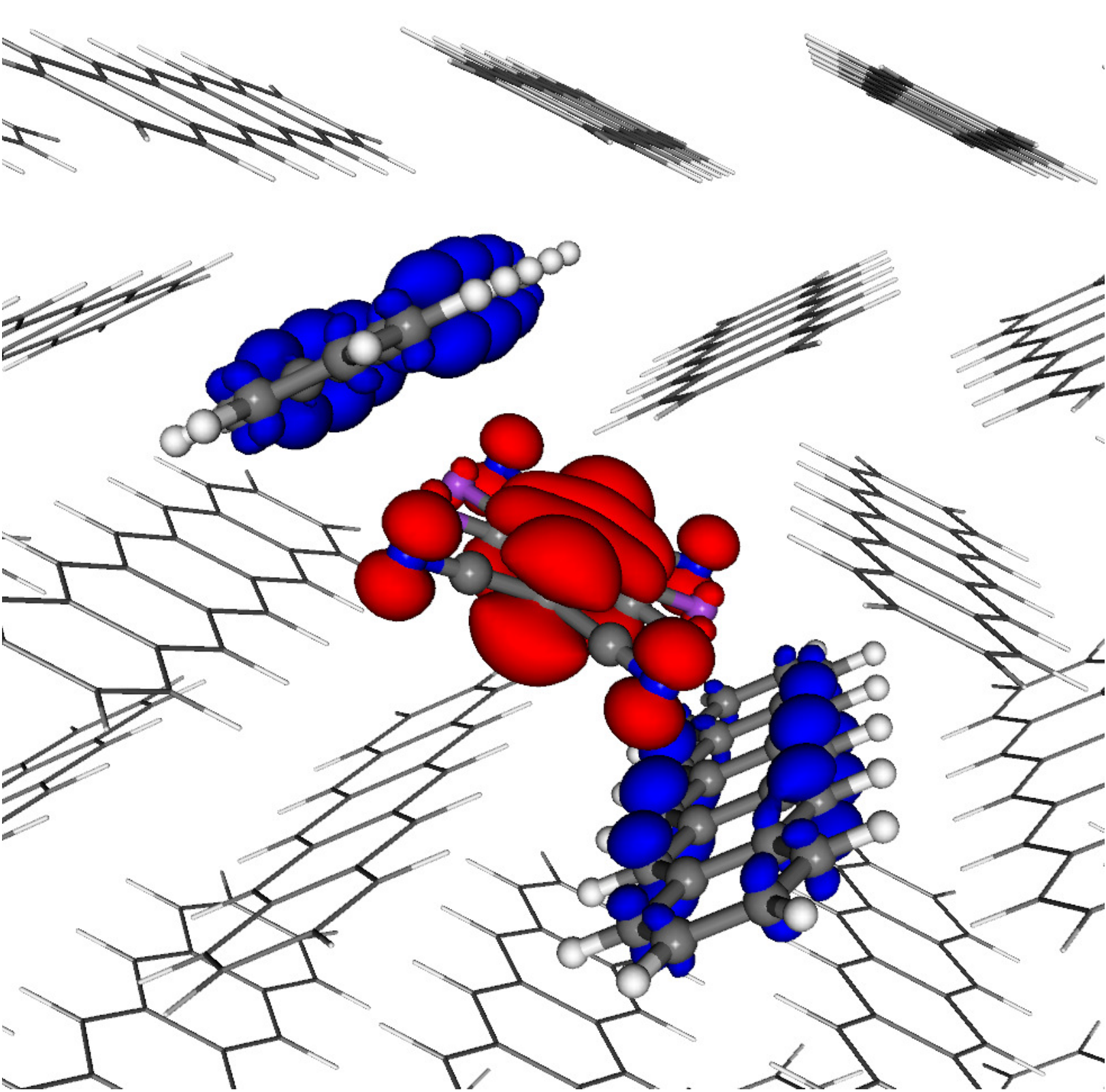}
\caption{(Color online)
Representation of the 1+2 CPX (QM region, ball-and-stick model) embedded in the pentacene crystals (MM region, wideframe representation) employed in the present validation.
The isocontour representation shows the averaged electron-hole density for the lowest Bethe-Salpeter $S_1$ excitation, presenting charge transfer along the same crystallographic direction as for the 1+6 CPX $S_1$  excitation studied in Section \ref{s:results}. 
}
\label{str}
\end{figure}

To perform extensive benchmark calculations at a reasonable cost, we study a 
relatively small pentacene-F4TCNQ complex (CPX) including the dopant and two of its pentacene 
neighbors, named 1+2 CPX and shown in Figure~\ref{str}.
The actual choice of such a geometry is motivated by the calculations on the larger 1+6 CPX discussed in Section \ref{s:results}, presenting its lowest-energy excitation S1 localized on the two molecules of the 1+2 CPX (see Figure~\ref{str}).

We first address the stability of the ev$GW$ and Bethe-Salpeter results, namely the fact that the calculated gap and excitation energies hardly depend on the Kohn-Sham eigenstates 
chosen as input for the $GW$ calculations. As shown in several publications, \cite{Sini11,Atalla13}  the Kohn-Sham HOMO-LUMO gap of donor-acceptor complexes varies dramatically 
with the amount of exact exchange. This is confirmed in the present case (see Table~\ref{functional}, Kohn-Sham column) with a variation from 0.23 eV to 1.64 eV when increasing the 
amount of exact exchange from 10$\%$ to 55$\%$.
 However, this variation is significantly removed in the resulting QM/MM (ev$GW$+CR) gap and corresponding Bethe-Salpeter (BSE) excitations 
(see Table~\ref{functional}). Such a stability originates from the ev$GW$ scheme that self-consistently corrects  the occupied/unoccupied electronic energy levels,
dramatically removing the starting point dependence. 
We show further that the PBEh($\alpha=0.4$) global hybrid functional (40$\%$ of exact exchange) Kohn-Sham eigenvalues best reproduces the  ev$GW$ gap, 
justifying the choice of the PBEh($\alpha=0.4$) Kohn-Sham eigenstates to start our 1+6 CPX calculations.

\begin{table}
\begin{tabular}{|c | c | c | c|}
\hline 
&\multicolumn{2}{|c|}{HOMO-LUMO gap} & BSE/ev$GW$+CR \\                      
\hline 
$\alpha$ & Kohn-Sham & ev$GW$+CR & S1 / S3 \\
\hline 
0.10 & 0.23 & 0.93 & 0.27/1.51 \\
0.25 & 0.58 & 0.85 & 0.11/1.43 \\
0.40 & 0.92 & 0.91 & 0.12/1.54 \\
0.55 & 1.64 & 0.93 & 0.16/1.59 \\
\hline 
\end{tabular}
\caption{Kohn-Sham and QM/MM (ev$GW$+CR) HOMO-LUMO gap and related Bethe-Salpeter lowest bright excitation energies as a function of the starting exchange-correlation PBEh($\alpha$) functional. \cite{Per96}
The parameter $\alpha$ gives the fraction of exact exchange. 
Calculations are performed for the 1+2 CPX  at the cc-pVTZ level.
Energies are in eV. 
For BSE excitation energies (Tamm-Dancoff approximation), we focus on the bright excitations originating from the HOMO and HOMO$-1$ pentacene subbands to the F4TCNQ LUMO. }
\label{functional}
\end{table}

We show in Table~\ref{basis} that the 6-311G(d) atomic basis set provides energy differences, namely HOMO-LUMO gap and excitation energies, well converged as compared to the ones obtained with the much larger correlation-consistent cc-pVTZ basis set.

\begin{table}[h]
\begin{tabular}{|c | c | c | c|}
\hline 
&\multicolumn{2}{|c|}{HOMO-LUMO gap} & BSE/ev$GW$+CR \\                      
\hline 
basis & Kohn-Sham & ev$GW$+CR  & S1 / S3 \\
\hline 
6-311G(d) & 0.89 & 0.93 & 0.15/1.55 \\
cc-pVTZ & 0.92 & 0.91 & 0.12/1.54 \\
\hline 
\end{tabular}
\caption{Kohn-Sham and QM/MM ev$GW$@PBEh(0.4) HOMO-LUMO gap and resulting BSE lowest bright singlet excitation energies as a function of the chosen basis
for the  1+2 CPX. Energies are in eV.}
\label{basis}
\end{table}
 
\clearpage
\subsection{Full BSE versus Tamm-Dancoff approximation and BSE Hamiltonian size}
We show in Table~\ref{BSE} that the Tamm-Dancoff approximation (TDA) provides charge-transfer (CT) excitation energies in excellent agreement with the full Bethe-Salpeter calculations that includes non-resonant excitations. Such a good agreement is expected for charge-transfer (CT) excitations for which the non-resonant matrix elements vanish. The first Frenkel excitation related to the onset of absorption of pentacene  is found to be blue-shifted by about 300 meV as expected. We further test the size of the (occupied)$\times$(unoccupied) product space used to set up the BSE electron-hole Hamiltonian matrix, namely the number of occupied/unoccupied levels included to build the BSE excitations. We compare calculations including occupied/unoccupied states located within 20 eV of the HOMO/LUMO, versus calculations including occupied/unoccupied states located within 10 eV of the HOMO/LUMO, as in our calculations on the 1+6 CPX. As expected, CT excitations, with much weight on the (HOMO-LUMO) transition and (HOMO$-1$)-LUMO for the excitations at 1.54-1.55 eV, are very well converged with the 10 eV cutoff. In contrast, the lowest pentacene-related transition with strong Frenkel character exhibits a slower convergence since it mixes higher energy levels.  

\begin{table}[h]
\begin{tabular}{|c | c | c |}
\hline & \multicolumn{2}{|c|}{BSE/ev$GW$+CR (QM/MM)} \\
\hline & S1 / S2 / S3 /S4 (CT) & pentacene (Frenkel-like) \\                  
\hline 
TDA  (20 eV) & 0.12/0.14/1.54/1.55 & 2.23 \\
Full (20 eV) & 0.12/0.13/1.54/1.55 & 1.93   \\
Full (10 eV) & 0.12/0.13/1.55/1.56 & 2.06 \\
\hline 
\end{tabular}
\caption{Comparison of BSE excitation energies in the 1+2 CPX system within the TDA and "full" approach as a function of the energy cutoff on the occupied/unoccupied states.
Energies are in eV.
}
\label{BSE}
\end{table}

\clearpage

\begin{acknowledgments}
This project has received funding from the European Union Horizon 2020 research and innovation programme under grant agreement No 646176 (EXTMOS). 
G.D. thanks Georg Heimel, Norbert Koch and Ingo Salzmann for stimulating discussions and acknowledges support from EU through the FP7-PEOPLE-2013-IEF program (GA 2013-625198).
D.B. is FNRS research director.
This research used computational resources from the French GENCI-CINES/IDRIS and the Belgian CECI/CENAERO. 
\end{acknowledgments}

\bibliographystyle{apsrev4-1}
\bibliography{doping} 

\begin{thebibliography}{55}%
\makeatletter
\providecommand \@ifxundefined [1]{%
 \@ifx{#1\undefined}
}%
\providecommand \@ifnum [1]{%
 \ifnum #1\expandafter \@firstoftwo
 \else \expandafter \@secondoftwo
 \fi
}%
\providecommand \@ifx [1]{%
 \ifx #1\expandafter \@firstoftwo
 \else \expandafter \@secondoftwo
 \fi
}%
\providecommand \natexlab [1]{#1}%
\providecommand \enquote  [1]{``#1''}%
\providecommand \bibnamefont  [1]{#1}%
\providecommand \bibfnamefont [1]{#1}%
\providecommand \citenamefont [1]{#1}%
\providecommand \href@noop [0]{\@secondoftwo}%
\providecommand \href [0]{\begingroup \@sanitize@url \@href}%
\providecommand \@href[1]{\@@startlink{#1}\@@href}%
\providecommand \@@href[1]{\endgroup#1\@@endlink}%
\providecommand \@sanitize@url [0]{\catcode `\\12\catcode `\$12\catcode
  `\&12\catcode `\#12\catcode `\^12\catcode `\_12\catcode `\%12\relax}%
\providecommand \@@startlink[1]{}%
\providecommand \@@endlink[0]{}%
\providecommand \url  [0]{\begingroup\@sanitize@url \@url }%
\providecommand \@url [1]{\endgroup\@href {#1}{\urlprefix }}%
\providecommand \urlprefix  [0]{URL }%
\providecommand \Eprint [0]{\href }%
\providecommand \doibase [0]{http://dx.doi.org/}%
\providecommand \selectlanguage [0]{\@gobble}%
\providecommand \bibinfo  [0]{\@secondoftwo}%
\providecommand \bibfield  [0]{\@secondoftwo}%
\providecommand \translation [1]{[#1]}%
\providecommand \BibitemOpen [0]{}%
\providecommand \bibitemStop [0]{}%
\providecommand \bibitemNoStop [0]{.\EOS\space}%
\providecommand \EOS [0]{\spacefactor3000\relax}%
\providecommand \BibitemShut  [1]{\csname bibitem#1\endcsname}%
\let\auto@bib@innerbib\@empty
\bibitem [{\citenamefont {Walzer}\ \emph {et~al.}(2007)\citenamefont {Walzer},
  \citenamefont {Maennig}, \citenamefont {Pfeiffer},\ and\ \citenamefont
  {Leo}}]{Wal07}%
  \BibitemOpen
  \bibfield  {author} {\bibinfo {author} {\bibfnamefont {K.}~\bibnamefont
  {Walzer}}, \bibinfo {author} {\bibfnamefont {B.}~\bibnamefont {Maennig}},
  \bibinfo {author} {\bibfnamefont {M.}~\bibnamefont {Pfeiffer}}, \ and\
  \bibinfo {author} {\bibfnamefont {K.}~\bibnamefont {Leo}},\ }\href {\doibase
  10.1021/cr050156n} {\bibfield  {journal} {\bibinfo  {journal} {Chem. Rev.}\
  }\textbf {\bibinfo {volume} {107}},\ \bibinfo {pages} {1233} (\bibinfo {year}
  {2007})}\BibitemShut {NoStop}%
\bibitem [{\citenamefont {Salzmann}\ \emph {et~al.}(2016)\citenamefont
  {Salzmann}, \citenamefont {Heimel}, \citenamefont {Oehzelt}, \citenamefont
  {Winkler},\ and\ \citenamefont {Koch}}]{Sal16}%
  \BibitemOpen
  \bibfield  {author} {\bibinfo {author} {\bibfnamefont {I.}~\bibnamefont
  {Salzmann}}, \bibinfo {author} {\bibfnamefont {G.}~\bibnamefont {Heimel}},
  \bibinfo {author} {\bibfnamefont {M.}~\bibnamefont {Oehzelt}}, \bibinfo
  {author} {\bibfnamefont {S.}~\bibnamefont {Winkler}}, \ and\ \bibinfo
  {author} {\bibfnamefont {N.}~\bibnamefont {Koch}},\ }\href {\doibase
  10.1021/acs.accounts.5b00438} {\bibfield  {journal} {\bibinfo  {journal}
  {Acc. Chem. Res.}\ }\textbf {\bibinfo {volume} {49}},\ \bibinfo {pages} {370}
  (\bibinfo {year} {2016})}\BibitemShut {NoStop}%
\bibitem [{\citenamefont {Sze}\ and\ \citenamefont {Ng}(2006)}]{Sze}%
  \BibitemOpen
  \bibfield  {author} {\bibinfo {author} {\bibfnamefont {S.~M.}\ \bibnamefont
  {Sze}}\ and\ \bibinfo {author} {\bibfnamefont {K.~K.}\ \bibnamefont {Ng}},\
  }\href@noop {} {\emph {\bibinfo {title} {Physics of Semiconductor Devices}}}\
  (\bibinfo  {publisher} {John Wiley and Sons},\ \bibinfo {year}
  {2006})\BibitemShut {NoStop}%
\bibitem [{\citenamefont {Salzmann}\ \emph {et~al.}(2012)\citenamefont
  {Salzmann}, \citenamefont {Heimel}, \citenamefont {Duhm}, \citenamefont
  {Oehzelt}, \citenamefont {Pingel}, \citenamefont {George}, \citenamefont
  {Schnegg}, \citenamefont {Lips}, \citenamefont {Blum}, \citenamefont
  {Vollmer},\ and\ \citenamefont {Koch}}]{Sal12}%
  \BibitemOpen
  \bibfield  {author} {\bibinfo {author} {\bibfnamefont {I.}~\bibnamefont
  {Salzmann}}, \bibinfo {author} {\bibfnamefont {G.}~\bibnamefont {Heimel}},
  \bibinfo {author} {\bibfnamefont {S.}~\bibnamefont {Duhm}}, \bibinfo {author}
  {\bibfnamefont {M.}~\bibnamefont {Oehzelt}}, \bibinfo {author} {\bibfnamefont
  {P.}~\bibnamefont {Pingel}}, \bibinfo {author} {\bibfnamefont {B.~M.}\
  \bibnamefont {George}}, \bibinfo {author} {\bibfnamefont {A.}~\bibnamefont
  {Schnegg}}, \bibinfo {author} {\bibfnamefont {K.}~\bibnamefont {Lips}},
  \bibinfo {author} {\bibfnamefont {R.-P.}\ \bibnamefont {Blum}}, \bibinfo
  {author} {\bibfnamefont {A.}~\bibnamefont {Vollmer}}, \ and\ \bibinfo
  {author} {\bibfnamefont {N.}~\bibnamefont {Koch}},\ }\href {\doibase
  10.1103/PhysRevLett.108.035502} {\bibfield  {journal} {\bibinfo  {journal}
  {Phys. Rev. Lett.}\ }\textbf {\bibinfo {volume} {108}},\ \bibinfo {pages}
  {035502} (\bibinfo {year} {2012})}\BibitemShut {NoStop}%
\bibitem [{\citenamefont {Duong}\ \emph {et~al.}(2013)\citenamefont {Duong},
  \citenamefont {Wang}, \citenamefont {Antono}, \citenamefont {Toney},\ and\
  \citenamefont {Salleo}}]{Duo13}%
  \BibitemOpen
  \bibfield  {author} {\bibinfo {author} {\bibfnamefont {D.~T.}\ \bibnamefont
  {Duong}}, \bibinfo {author} {\bibfnamefont {C.}~\bibnamefont {Wang}},
  \bibinfo {author} {\bibfnamefont {E.}~\bibnamefont {Antono}}, \bibinfo
  {author} {\bibfnamefont {M.~F.}\ \bibnamefont {Toney}}, \ and\ \bibinfo
  {author} {\bibfnamefont {A.}~\bibnamefont {Salleo}},\ }\href {\doibase
  10.1016/j.orgel.2013.02.028} {\bibfield  {journal} {\bibinfo  {journal} {Org.
  Electron.}\ }\textbf {\bibinfo {volume} {14}},\ \bibinfo {pages} {1330 }
  (\bibinfo {year} {2013})}\BibitemShut {NoStop}%
\bibitem [{\citenamefont {Pingel}\ and\ \citenamefont {Neher}(2013)}]{Pin13}%
  \BibitemOpen
  \bibfield  {author} {\bibinfo {author} {\bibfnamefont {P.}~\bibnamefont
  {Pingel}}\ and\ \bibinfo {author} {\bibfnamefont {D.}~\bibnamefont {Neher}},\
  }\href {\doibase 10.1103/PhysRevB.87.115209} {\bibfield  {journal} {\bibinfo
  {journal} {Phys. Rev. B}\ }\textbf {\bibinfo {volume} {87}},\ \bibinfo
  {pages} {115209} (\bibinfo {year} {2013})}\BibitemShut {NoStop}%
\bibitem [{\citenamefont {Mityashin}\ \emph {et~al.}(2012)\citenamefont
  {Mityashin}, \citenamefont {Olivier}, \citenamefont {Van~Regemorter},
  \citenamefont {Rolin}, \citenamefont {Verlaak}, \citenamefont {Martinelli},
  \citenamefont {Beljonne}, \citenamefont {Cornil}, \citenamefont {Genoe},\
  and\ \citenamefont {Heremans}}]{Mit12}%
  \BibitemOpen
  \bibfield  {author} {\bibinfo {author} {\bibfnamefont {A.}~\bibnamefont
  {Mityashin}}, \bibinfo {author} {\bibfnamefont {Y.}~\bibnamefont {Olivier}},
  \bibinfo {author} {\bibfnamefont {T.}~\bibnamefont {Van~Regemorter}},
  \bibinfo {author} {\bibfnamefont {C.}~\bibnamefont {Rolin}}, \bibinfo
  {author} {\bibfnamefont {S.}~\bibnamefont {Verlaak}}, \bibinfo {author}
  {\bibfnamefont {N.~G.}\ \bibnamefont {Martinelli}}, \bibinfo {author}
  {\bibfnamefont {D.}~\bibnamefont {Beljonne}}, \bibinfo {author}
  {\bibfnamefont {J.}~\bibnamefont {Cornil}}, \bibinfo {author} {\bibfnamefont
  {J.}~\bibnamefont {Genoe}}, \ and\ \bibinfo {author} {\bibfnamefont
  {P.}~\bibnamefont {Heremans}},\ }\href {\doibase 10.1002/adma.201104269}
  {\bibfield  {journal} {\bibinfo  {journal} {Adv. Mater.}\ }\textbf {\bibinfo
  {volume} {24}},\ \bibinfo {pages} {1535} (\bibinfo {year}
  {2012})}\BibitemShut {NoStop}%
\bibitem [{\citenamefont {M{\'e}ndez}\ \emph {et~al.}(2015)\citenamefont
  {M{\'e}ndez}, \citenamefont {Heimel}, \citenamefont {Winkler}, \citenamefont
  {Frisch}, \citenamefont {Opitz}, \citenamefont {Sauer}, \citenamefont
  {Wegner}, \citenamefont {Oehzelt}, \citenamefont {R{\"o}thel}, \citenamefont
  {Duhm}, \citenamefont {T{\"o}bbens}, \citenamefont {Koch},\ and\
  \citenamefont {Salzmann}}]{Men15}%
  \BibitemOpen
  \bibfield  {author} {\bibinfo {author} {\bibfnamefont {H.}~\bibnamefont
  {M{\'e}ndez}}, \bibinfo {author} {\bibfnamefont {G.}~\bibnamefont {Heimel}},
  \bibinfo {author} {\bibfnamefont {S.}~\bibnamefont {Winkler}}, \bibinfo
  {author} {\bibfnamefont {J.}~\bibnamefont {Frisch}}, \bibinfo {author}
  {\bibfnamefont {A.}~\bibnamefont {Opitz}}, \bibinfo {author} {\bibfnamefont
  {K.}~\bibnamefont {Sauer}}, \bibinfo {author} {\bibfnamefont
  {B.}~\bibnamefont {Wegner}}, \bibinfo {author} {\bibfnamefont
  {M.}~\bibnamefont {Oehzelt}}, \bibinfo {author} {\bibfnamefont
  {C.}~\bibnamefont {R{\"o}thel}}, \bibinfo {author} {\bibfnamefont
  {S.}~\bibnamefont {Duhm}}, \bibinfo {author} {\bibfnamefont {D.}~\bibnamefont
  {T{\"o}bbens}}, \bibinfo {author} {\bibfnamefont {N.}~\bibnamefont {Koch}}, \
  and\ \bibinfo {author} {\bibfnamefont {I.}~\bibnamefont {Salzmann}},\ }\href
  {\doibase 10.1038/ncomms9560} {\bibfield  {journal} {\bibinfo  {journal}
  {Nat. Commun.}\ }\textbf {\bibinfo {volume} {6}},\ \bibinfo {pages} {8560}
  (\bibinfo {year} {2015})}\BibitemShut {NoStop}%
\bibitem [{\citenamefont {Png}\ \emph {et~al.}(2016)\citenamefont {Png},
  \citenamefont {Ang}, \citenamefont {Teo}, \citenamefont {Choo}, \citenamefont
  {Tang}, \citenamefont {Belaineh}, \citenamefont {Chua},\ and\ \citenamefont
  {Ho}}]{Png16}%
  \BibitemOpen
  \bibfield  {author} {\bibinfo {author} {\bibfnamefont {R.-Q.}\ \bibnamefont
  {Png}}, \bibinfo {author} {\bibfnamefont {M.~C.~Y.}\ \bibnamefont {Ang}},
  \bibinfo {author} {\bibfnamefont {M.-H.}\ \bibnamefont {Teo}}, \bibinfo
  {author} {\bibfnamefont {K.-K.}\ \bibnamefont {Choo}}, \bibinfo {author}
  {\bibfnamefont {C.~G.}\ \bibnamefont {Tang}}, \bibinfo {author}
  {\bibfnamefont {D.}~\bibnamefont {Belaineh}}, \bibinfo {author}
  {\bibfnamefont {L.-L.}\ \bibnamefont {Chua}}, \ and\ \bibinfo {author}
  {\bibfnamefont {P.~K.~H.}\ \bibnamefont {Ho}},\ }\href@noop {} {\bibfield
  {journal} {\bibinfo  {journal} {Nat. Commun.}\ }\textbf {\bibinfo {volume}
  {7}},\ \bibinfo {pages} {11948} (\bibinfo {year} {2016})}\BibitemShut
  {NoStop}%
\bibitem [{\citenamefont {Karpov}\ \emph {et~al.}(2016)\citenamefont {Karpov},
  \citenamefont {Erdmann}, \citenamefont {Raguzin}, \citenamefont {Al-Hussein},
  \citenamefont {Binner}, \citenamefont {Lappan}, \citenamefont {Stamm},
  \citenamefont {Gerasimov}, \citenamefont {Beryozkina}, \citenamefont
  {Bakulev}, \citenamefont {Anokhin}, \citenamefont {Ivanov}, \citenamefont
  {Günther}, \citenamefont {Gemming}, \citenamefont {Seifert}, \citenamefont
  {Voit}, \citenamefont {Di~Pietro},\ and\ \citenamefont {Kiriy}}]{Kar16}%
  \BibitemOpen
  \bibfield  {author} {\bibinfo {author} {\bibfnamefont {Y.}~\bibnamefont
  {Karpov}}, \bibinfo {author} {\bibfnamefont {T.}~\bibnamefont {Erdmann}},
  \bibinfo {author} {\bibfnamefont {I.}~\bibnamefont {Raguzin}}, \bibinfo
  {author} {\bibfnamefont {M.}~\bibnamefont {Al-Hussein}}, \bibinfo {author}
  {\bibfnamefont {M.}~\bibnamefont {Binner}}, \bibinfo {author} {\bibfnamefont
  {U.}~\bibnamefont {Lappan}}, \bibinfo {author} {\bibfnamefont
  {M.}~\bibnamefont {Stamm}}, \bibinfo {author} {\bibfnamefont {K.~L.}\
  \bibnamefont {Gerasimov}}, \bibinfo {author} {\bibfnamefont {T.}~\bibnamefont
  {Beryozkina}}, \bibinfo {author} {\bibfnamefont {V.}~\bibnamefont {Bakulev}},
  \bibinfo {author} {\bibfnamefont {D.~V.}\ \bibnamefont {Anokhin}}, \bibinfo
  {author} {\bibfnamefont {D.~A.}\ \bibnamefont {Ivanov}}, \bibinfo {author}
  {\bibfnamefont {F.}~\bibnamefont {Günther}}, \bibinfo {author}
  {\bibfnamefont {S.}~\bibnamefont {Gemming}}, \bibinfo {author} {\bibfnamefont
  {G.}~\bibnamefont {Seifert}}, \bibinfo {author} {\bibfnamefont
  {B.}~\bibnamefont {Voit}}, \bibinfo {author} {\bibfnamefont {R.}~\bibnamefont
  {Di~Pietro}}, \ and\ \bibinfo {author} {\bibfnamefont {A.}~\bibnamefont
  {Kiriy}},\ }\href {\doibase 10.1002/adma.201506295} {\bibfield  {journal}
  {\bibinfo  {journal} {Adv. Mater.}\ }\textbf {\bibinfo {volume} {28}},\
  \bibinfo {pages} {6003} (\bibinfo {year} {2016})}\BibitemShut {NoStop}%
\bibitem [{\citenamefont {Kang}\ \emph {et~al.}(2016)\citenamefont {Kang},
  \citenamefont {Watanabe}, \citenamefont {Broch}, \citenamefont {Sepe},
  \citenamefont {Brown}, \citenamefont {Nasrallah}, \citenamefont {Nikolka},
  \citenamefont {Fei}, \citenamefont {Heeney}, \citenamefont {Matsumoto},
  \citenamefont {Marumoto}, \citenamefont {Tanaka}, \citenamefont {Kuroda},\
  and\ \citenamefont {Sirringhaus}}]{Kang16}%
  \BibitemOpen
  \bibfield  {author} {\bibinfo {author} {\bibfnamefont {K.}~\bibnamefont
  {Kang}}, \bibinfo {author} {\bibfnamefont {S.}~\bibnamefont {Watanabe}},
  \bibinfo {author} {\bibfnamefont {K.}~\bibnamefont {Broch}}, \bibinfo
  {author} {\bibfnamefont {A.}~\bibnamefont {Sepe}}, \bibinfo {author}
  {\bibfnamefont {A.}~\bibnamefont {Brown}}, \bibinfo {author} {\bibfnamefont
  {I.}~\bibnamefont {Nasrallah}}, \bibinfo {author} {\bibfnamefont
  {M.}~\bibnamefont {Nikolka}}, \bibinfo {author} {\bibfnamefont
  {Z.}~\bibnamefont {Fei}}, \bibinfo {author} {\bibfnamefont {M.}~\bibnamefont
  {Heeney}}, \bibinfo {author} {\bibfnamefont {D.}~\bibnamefont {Matsumoto}},
  \bibinfo {author} {\bibfnamefont {K.}~\bibnamefont {Marumoto}}, \bibinfo
  {author} {\bibfnamefont {H.}~\bibnamefont {Tanaka}}, \bibinfo {author}
  {\bibfnamefont {S.-i.}\ \bibnamefont {Kuroda}}, \ and\ \bibinfo {author}
  {\bibfnamefont {H.}~\bibnamefont {Sirringhaus}},\ }\href {\doibase
  10.1038/nmat4634} {\bibfield  {journal} {\bibinfo  {journal} {Nat. Mater.}\
  }\textbf {\bibinfo {volume} {15}},\ \bibinfo {pages} {896} (\bibinfo {year}
  {2016})}\BibitemShut {NoStop}%
\bibitem [{\citenamefont {Pingel}\ \emph {et~al.}(2010)\citenamefont {Pingel},
  \citenamefont {Zhu}, \citenamefont {Park}, \citenamefont {Vogel},
  \citenamefont {Janietz}, \citenamefont {Kim}, \citenamefont {Rabe},
  \citenamefont {Brédas},\ and\ \citenamefont {Koch}}]{Pin10}%
  \BibitemOpen
  \bibfield  {author} {\bibinfo {author} {\bibfnamefont {P.}~\bibnamefont
  {Pingel}}, \bibinfo {author} {\bibfnamefont {L.}~\bibnamefont {Zhu}},
  \bibinfo {author} {\bibfnamefont {K.~S.}\ \bibnamefont {Park}}, \bibinfo
  {author} {\bibfnamefont {J.-O.}\ \bibnamefont {Vogel}}, \bibinfo {author}
  {\bibfnamefont {S.}~\bibnamefont {Janietz}}, \bibinfo {author} {\bibfnamefont
  {E.-G.}\ \bibnamefont {Kim}}, \bibinfo {author} {\bibfnamefont {J.~P.}\
  \bibnamefont {Rabe}}, \bibinfo {author} {\bibfnamefont {J.-L.}\ \bibnamefont
  {Brédas}}, \ and\ \bibinfo {author} {\bibfnamefont {N.}~\bibnamefont
  {Koch}},\ }\href {\doibase 10.1021/jz100492c} {\bibfield  {journal} {\bibinfo
   {journal} {Chem. Phys. Lett.}\ }\textbf {\bibinfo {volume} {1}},\ \bibinfo
  {pages} {2037} (\bibinfo {year} {2010})}\BibitemShut {NoStop}%
\bibitem [{\citenamefont {Ghani}\ \emph {et~al.}(2015)\citenamefont {Ghani},
  \citenamefont {Opitz}, \citenamefont {Pingel}, \citenamefont {Heimel},
  \citenamefont {Salzmann}, \citenamefont {Frisch}, \citenamefont {Neher},
  \citenamefont {Tsami}, \citenamefont {Scherf},\ and\ \citenamefont
  {Koch}}]{Gha15}%
  \BibitemOpen
  \bibfield  {author} {\bibinfo {author} {\bibfnamefont {F.}~\bibnamefont
  {Ghani}}, \bibinfo {author} {\bibfnamefont {A.}~\bibnamefont {Opitz}},
  \bibinfo {author} {\bibfnamefont {P.}~\bibnamefont {Pingel}}, \bibinfo
  {author} {\bibfnamefont {G.}~\bibnamefont {Heimel}}, \bibinfo {author}
  {\bibfnamefont {I.}~\bibnamefont {Salzmann}}, \bibinfo {author}
  {\bibfnamefont {J.}~\bibnamefont {Frisch}}, \bibinfo {author} {\bibfnamefont
  {D.}~\bibnamefont {Neher}}, \bibinfo {author} {\bibfnamefont
  {A.}~\bibnamefont {Tsami}}, \bibinfo {author} {\bibfnamefont
  {U.}~\bibnamefont {Scherf}}, \ and\ \bibinfo {author} {\bibfnamefont
  {N.}~\bibnamefont {Koch}},\ }\href {\doibase 10.1002/polb.23631} {\bibfield
  {journal} {\bibinfo  {journal} {J. Polym. Sci., Part B: Polym. Phys.}\
  }\textbf {\bibinfo {volume} {53}},\ \bibinfo {pages} {58} (\bibinfo {year}
  {2015})}\BibitemShut {NoStop}%
\bibitem [{\citenamefont {M\'{e}ndez}\ \emph {et~al.}(2013)\citenamefont
  {M\'{e}ndez}, \citenamefont {Heimel}, \citenamefont {Opitz}, \citenamefont
  {Sauer}, \citenamefont {Barkowski}, \citenamefont {Oehzelt}, \citenamefont
  {Soeda}, \citenamefont {Okamoto}, \citenamefont {Takeya}, \citenamefont
  {Arlin}, \citenamefont {Balandier}, \citenamefont {Geerts}, \citenamefont
  {Koch},\ and\ \citenamefont {Salzmann}}]{Men13}%
  \BibitemOpen
  \bibfield  {author} {\bibinfo {author} {\bibfnamefont {H.}~\bibnamefont
  {M\'{e}ndez}}, \bibinfo {author} {\bibfnamefont {G.}~\bibnamefont {Heimel}},
  \bibinfo {author} {\bibfnamefont {A.}~\bibnamefont {Opitz}}, \bibinfo
  {author} {\bibfnamefont {K.}~\bibnamefont {Sauer}}, \bibinfo {author}
  {\bibfnamefont {P.}~\bibnamefont {Barkowski}}, \bibinfo {author}
  {\bibfnamefont {M.}~\bibnamefont {Oehzelt}}, \bibinfo {author} {\bibfnamefont
  {J.}~\bibnamefont {Soeda}}, \bibinfo {author} {\bibfnamefont
  {T.}~\bibnamefont {Okamoto}}, \bibinfo {author} {\bibfnamefont
  {J.}~\bibnamefont {Takeya}}, \bibinfo {author} {\bibfnamefont {J.-B.}\
  \bibnamefont {Arlin}}, \bibinfo {author} {\bibfnamefont {J.-Y.}\ \bibnamefont
  {Balandier}}, \bibinfo {author} {\bibfnamefont {Y.}~\bibnamefont {Geerts}},
  \bibinfo {author} {\bibfnamefont {N.}~\bibnamefont {Koch}}, \ and\ \bibinfo
  {author} {\bibfnamefont {I.}~\bibnamefont {Salzmann}},\ }\href {\doibase
  10.1002/anie.201302396} {\bibfield  {journal} {\bibinfo  {journal} {Angew.
  Chem.-Int. Edit.}\ }\textbf {\bibinfo {volume} {52}},\ \bibinfo {pages}
  {7751} (\bibinfo {year} {2013})}\BibitemShut {NoStop}%
\bibitem [{\citenamefont {Ha}\ and\ \citenamefont {Kahn}(2009)}]{Ha09}%
  \BibitemOpen
  \bibfield  {author} {\bibinfo {author} {\bibfnamefont {S.~D.}\ \bibnamefont
  {Ha}}\ and\ \bibinfo {author} {\bibfnamefont {A.}~\bibnamefont {Kahn}},\
  }\href {\doibase 10.1103/PhysRevB.80.195410} {\bibfield  {journal} {\bibinfo
  {journal} {Phys. Rev. B}\ }\textbf {\bibinfo {volume} {80}},\ \bibinfo
  {pages} {195410} (\bibinfo {year} {2009})}\BibitemShut {NoStop}%
\bibitem [{\citenamefont {Kleemann}\ \emph {et~al.}(2012)\citenamefont
  {Kleemann}, \citenamefont {Schuenemann}, \citenamefont {Zakhidov},
  \citenamefont {Riede}, \citenamefont {Lüssem},\ and\ \citenamefont
  {Leo}}]{Kle12}%
  \BibitemOpen
  \bibfield  {author} {\bibinfo {author} {\bibfnamefont {H.}~\bibnamefont
  {Kleemann}}, \bibinfo {author} {\bibfnamefont {C.}~\bibnamefont
  {Schuenemann}}, \bibinfo {author} {\bibfnamefont {A.~A.}\ \bibnamefont
  {Zakhidov}}, \bibinfo {author} {\bibfnamefont {M.}~\bibnamefont {Riede}},
  \bibinfo {author} {\bibfnamefont {B.}~\bibnamefont {Lüssem}}, \ and\
  \bibinfo {author} {\bibfnamefont {K.}~\bibnamefont {Leo}},\ }\href {\doibase
  10.1016/j.orgel.2011.09.027} {\bibfield  {journal} {\bibinfo  {journal} {Org.
  Electron.}\ }\textbf {\bibinfo {volume} {13}},\ \bibinfo {pages} {58 }
  (\bibinfo {year} {2012})}\BibitemShut {NoStop}%
\bibitem [{\citenamefont {Bredas}\ and\ \citenamefont {Street}(1985)}]{Bre85}%
  \BibitemOpen
  \bibfield  {author} {\bibinfo {author} {\bibfnamefont {J.~L.}\ \bibnamefont
  {Bredas}}\ and\ \bibinfo {author} {\bibfnamefont {G.~B.}\ \bibnamefont
  {Street}},\ }\href {\doibase 10.1021/ar00118a005} {\bibfield  {journal}
  {\bibinfo  {journal} {Acc. Chem. Res.}\ }\textbf {\bibinfo {volume} {18}},\
  \bibinfo {pages} {309} (\bibinfo {year} {1985})}\BibitemShut {NoStop}%
\bibitem [{\citenamefont {Topham}\ and\ \citenamefont {Soos}(2011)}]{Top11}%
  \BibitemOpen
  \bibfield  {author} {\bibinfo {author} {\bibfnamefont {B.~J.}\ \bibnamefont
  {Topham}}\ and\ \bibinfo {author} {\bibfnamefont {Z.~G.}\ \bibnamefont
  {Soos}},\ }\href {\doibase 10.1103/PhysRevB.84.165405} {\bibfield  {journal}
  {\bibinfo  {journal} {Phys. Rev. B}\ }\textbf {\bibinfo {volume} {84}},\
  \bibinfo {pages} {165405} (\bibinfo {year} {2011})}\BibitemShut {NoStop}%
\bibitem [{\citenamefont {Hedin}(1965)}]{Hed65}%
  \BibitemOpen
  \bibfield  {author} {\bibinfo {author} {\bibfnamefont {L.}~\bibnamefont
  {Hedin}},\ }\href {\doibase 10.1103/PhysRev.139.A796} {\bibfield  {journal}
  {\bibinfo  {journal} {Phys. Rev.}\ }\textbf {\bibinfo {volume} {139}},\
  \bibinfo {pages} {A796} (\bibinfo {year} {1965})}\BibitemShut {NoStop}%
\bibitem [{\citenamefont {Onida}\ \emph {et~al.}(2002)\citenamefont {Onida},
  \citenamefont {Reining},\ and\ \citenamefont {Rubio}}]{Oni02}%
  \BibitemOpen
  \bibfield  {author} {\bibinfo {author} {\bibfnamefont {G.}~\bibnamefont
  {Onida}}, \bibinfo {author} {\bibfnamefont {L.}~\bibnamefont {Reining}}, \
  and\ \bibinfo {author} {\bibfnamefont {A.}~\bibnamefont {Rubio}},\ }\href
  {\doibase 10.1103/RevModPhys.74.601} {\bibfield  {journal} {\bibinfo
  {journal} {Rev. Mod. Phys.}\ }\textbf {\bibinfo {volume} {74}},\ \bibinfo
  {pages} {601} (\bibinfo {year} {2002})}\BibitemShut {NoStop}%
\bibitem [{\citenamefont {Hanke}\ and\ \citenamefont {Sham}(1979)}]{Han79}%
  \BibitemOpen
  \bibfield  {author} {\bibinfo {author} {\bibfnamefont {W.}~\bibnamefont
  {Hanke}}\ and\ \bibinfo {author} {\bibfnamefont {L.~J.}\ \bibnamefont
  {Sham}},\ }\href {\doibase 10.1103/PhysRevLett.43.387} {\bibfield  {journal}
  {\bibinfo  {journal} {Phys. Rev. Lett.}\ }\textbf {\bibinfo {volume} {43}},\
  \bibinfo {pages} {387} (\bibinfo {year} {1979})}\BibitemShut {NoStop}%
\bibitem [{\citenamefont {Knight}\ \emph {et~al.}(2016)\citenamefont {Knight},
  \citenamefont {Wang}, \citenamefont {Gallandi}, \citenamefont
  {Dolgounitcheva}, \citenamefont {Ren}, \citenamefont {Ortiz}, \citenamefont
  {Rinke}, \citenamefont {K\"{o}rzd\"{o}rfer},\ and\ \citenamefont
  {Marom}}]{Kni16}%
  \BibitemOpen
  \bibfield  {author} {\bibinfo {author} {\bibfnamefont {J.~W.}\ \bibnamefont
  {Knight}}, \bibinfo {author} {\bibfnamefont {X.}~\bibnamefont {Wang}},
  \bibinfo {author} {\bibfnamefont {L.}~\bibnamefont {Gallandi}}, \bibinfo
  {author} {\bibfnamefont {O.}~\bibnamefont {Dolgounitcheva}}, \bibinfo
  {author} {\bibfnamefont {X.}~\bibnamefont {Ren}}, \bibinfo {author}
  {\bibfnamefont {J.~V.}\ \bibnamefont {Ortiz}}, \bibinfo {author}
  {\bibfnamefont {P.}~\bibnamefont {Rinke}}, \bibinfo {author} {\bibfnamefont
  {T.}~\bibnamefont {K\"{o}rzd\"{o}rfer}}, \ and\ \bibinfo {author}
  {\bibfnamefont {N.}~\bibnamefont {Marom}},\ }\href {\doibase
  10.1021/acs.jctc.5b00871} {\bibfield  {journal} {\bibinfo  {journal} {J.
  Chem. Theory Comput.}\ }\textbf {\bibinfo {volume} {12}},\ \bibinfo {pages}
  {615} (\bibinfo {year} {2016})}\BibitemShut {NoStop}%
\bibitem [{\citenamefont {Kaplan}\ \emph {et~al.}(2016)\citenamefont {Kaplan},
  \citenamefont {Harding}, \citenamefont {Seiler}, \citenamefont {Weigend},
  \citenamefont {Evers},\ and\ \citenamefont {van Setten}}]{Kap16}%
  \BibitemOpen
  \bibfield  {author} {\bibinfo {author} {\bibfnamefont {F.}~\bibnamefont
  {Kaplan}}, \bibinfo {author} {\bibfnamefont {M.~E.}\ \bibnamefont {Harding}},
  \bibinfo {author} {\bibfnamefont {C.}~\bibnamefont {Seiler}}, \bibinfo
  {author} {\bibfnamefont {F.}~\bibnamefont {Weigend}}, \bibinfo {author}
  {\bibfnamefont {F.}~\bibnamefont {Evers}}, \ and\ \bibinfo {author}
  {\bibfnamefont {M.~J.}\ \bibnamefont {van Setten}},\ }\href {\doibase
  10.1021/acs.jctc.5b01238} {\bibfield  {journal} {\bibinfo  {journal} {J.
  Chem. Theory Comput.}\ }\textbf {\bibinfo {volume} {12}},\ \bibinfo {pages}
  {2528} (\bibinfo {year} {2016})}\BibitemShut {NoStop}%
\bibitem [{\citenamefont {Rangel}\ \emph {et~al.}(2016)\citenamefont {Rangel},
  \citenamefont {Hamed}, \citenamefont {Bruneval},\ and\ \citenamefont
  {Neaton}}]{Ran16b}%
  \BibitemOpen
  \bibfield  {author} {\bibinfo {author} {\bibfnamefont {T.}~\bibnamefont
  {Rangel}}, \bibinfo {author} {\bibfnamefont {S.~M.}\ \bibnamefont {Hamed}},
  \bibinfo {author} {\bibfnamefont {F.}~\bibnamefont {Bruneval}}, \ and\
  \bibinfo {author} {\bibfnamefont {J.~B.}\ \bibnamefont {Neaton}},\ }\href
  {\doibase 10.1021/acs.jctc.6b00163} {\bibfield  {journal} {\bibinfo
  {journal} {J. Chem. Theory Comput.}\ }\textbf {\bibinfo {volume} {12}},\
  \bibinfo {pages} {2834 } (\bibinfo {year} {2016})}\BibitemShut {NoStop}%
\bibitem [{\citenamefont {Bruneval}\ \emph {et~al.}(2015)\citenamefont
  {Bruneval}, \citenamefont {Hamed},\ and\ \citenamefont {Neaton}}]{Bru15}%
  \BibitemOpen
  \bibfield  {author} {\bibinfo {author} {\bibfnamefont {F.}~\bibnamefont
  {Bruneval}}, \bibinfo {author} {\bibfnamefont {S.~M.}\ \bibnamefont {Hamed}},
  \ and\ \bibinfo {author} {\bibfnamefont {J.~B.}\ \bibnamefont {Neaton}},\
  }\href {\doibase 10.1063/1.4922489} {\bibfield  {journal} {\bibinfo
  {journal} {J. Chem. Phys.}\ }\textbf {\bibinfo {volume} {142}},\ \bibinfo
  {eid} {244101} (\bibinfo {year} {2015})}\BibitemShut {NoStop}%
\bibitem [{\citenamefont {Jacquemin}\ \emph {et~al.}(2015)\citenamefont
  {Jacquemin}, \citenamefont {Duchemin},\ and\ \citenamefont {Blase}}]{Jac15a}%
  \BibitemOpen
  \bibfield  {author} {\bibinfo {author} {\bibfnamefont {D.}~\bibnamefont
  {Jacquemin}}, \bibinfo {author} {\bibfnamefont {I.}~\bibnamefont {Duchemin}},
  \ and\ \bibinfo {author} {\bibfnamefont {X.}~\bibnamefont {Blase}},\ }\href
  {\doibase 10.1021/acs.jctc.5b00304} {\bibfield  {journal} {\bibinfo
  {journal} {J. Chem. Theory Comput.}\ }\textbf {\bibinfo {volume} {11}},\
  \bibinfo {pages} {3290} (\bibinfo {year} {2015})}\BibitemShut {NoStop}%
\bibitem [{\citenamefont {Baumeier}\ \emph {et~al.}(2012)\citenamefont
  {Baumeier}, \citenamefont {Andrienko},\ and\ \citenamefont
  {Rohlfing}}]{Bau12}%
  \BibitemOpen
  \bibfield  {author} {\bibinfo {author} {\bibfnamefont {B.}~\bibnamefont
  {Baumeier}}, \bibinfo {author} {\bibfnamefont {D.}~\bibnamefont {Andrienko}},
  \ and\ \bibinfo {author} {\bibfnamefont {M.}~\bibnamefont {Rohlfing}},\
  }\href {\doibase 10.1021/ct300311x} {\bibfield  {journal} {\bibinfo
  {journal} {J. Chem. Theory Comput.}\ }\textbf {\bibinfo {volume} {8}},\
  \bibinfo {pages} {2790} (\bibinfo {year} {2012})}\BibitemShut {NoStop}%
\bibitem [{\citenamefont {Duchemin}\ \emph {et~al.}(2012)\citenamefont
  {Duchemin}, \citenamefont {Deutsch},\ and\ \citenamefont {Blase}}]{Duc12}%
  \BibitemOpen
  \bibfield  {author} {\bibinfo {author} {\bibfnamefont {I.}~\bibnamefont
  {Duchemin}}, \bibinfo {author} {\bibfnamefont {T.}~\bibnamefont {Deutsch}}, \
  and\ \bibinfo {author} {\bibfnamefont {X.}~\bibnamefont {Blase}},\ }\href
  {\doibase 10.1103/PhysRevLett.109.167801} {\bibfield  {journal} {\bibinfo
  {journal} {Phys. Rev. Lett.}\ }\textbf {\bibinfo {volume} {109}},\ \bibinfo
  {pages} {167801} (\bibinfo {year} {2012})}\BibitemShut {NoStop}%
\bibitem [{\citenamefont {Li}\ \emph {et~al.}(2016)\citenamefont {Li},
  \citenamefont {D'Avino}, \citenamefont {Duchemin}, \citenamefont {Beljonne},\
  and\ \citenamefont {Blase}}]{Li16}%
  \BibitemOpen
  \bibfield  {author} {\bibinfo {author} {\bibfnamefont {J.}~\bibnamefont
  {Li}}, \bibinfo {author} {\bibfnamefont {G.}~\bibnamefont {D'Avino}},
  \bibinfo {author} {\bibfnamefont {I.}~\bibnamefont {Duchemin}}, \bibinfo
  {author} {\bibfnamefont {D.}~\bibnamefont {Beljonne}}, \ and\ \bibinfo
  {author} {\bibfnamefont {X.}~\bibnamefont {Blase}},\ }\href {\doibase
  10.1021/acs.jpclett.6b01302} {\bibfield  {journal} {\bibinfo  {journal} {J.
  Phys. Chem. Lett.}\ }\textbf {\bibinfo {volume} {7}},\ \bibinfo {pages}
  {2814} (\bibinfo {year} {2016})}\BibitemShut {NoStop}%
\bibitem [{\citenamefont {D’Avino}\ \emph
  {et~al.}(2016{\natexlab{a}})\citenamefont {D’Avino}, \citenamefont
  {Muccioli}, \citenamefont {Castet}, \citenamefont {Poelking}, \citenamefont
  {Andrienko}, \citenamefont {Soos}, \citenamefont {Cornil},\ and\
  \citenamefont {Beljonne}}]{Dav16rev}%
  \BibitemOpen
  \bibfield  {author} {\bibinfo {author} {\bibfnamefont {G.}~\bibnamefont
  {D’Avino}}, \bibinfo {author} {\bibfnamefont {L.}~\bibnamefont {Muccioli}},
  \bibinfo {author} {\bibfnamefont {F.}~\bibnamefont {Castet}}, \bibinfo
  {author} {\bibfnamefont {C.}~\bibnamefont {Poelking}}, \bibinfo {author}
  {\bibfnamefont {D.}~\bibnamefont {Andrienko}}, \bibinfo {author}
  {\bibfnamefont {Z.~G.}\ \bibnamefont {Soos}}, \bibinfo {author}
  {\bibfnamefont {J.}~\bibnamefont {Cornil}}, \ and\ \bibinfo {author}
  {\bibfnamefont {D.}~\bibnamefont {Beljonne}},\ }\href {\doibase
  10.1088/0953-8984/28/43/433002} {\bibfield  {journal} {\bibinfo  {journal}
  {J. Phys.: Condens. Matter}\ }\textbf {\bibinfo {volume} {28}},\ \bibinfo
  {pages} {433002} (\bibinfo {year} {2016}{\natexlab{a}})}\BibitemShut
  {NoStop}%
\bibitem [{\citenamefont {Blase}\ \emph {et~al.}(2011)\citenamefont {Blase},
  \citenamefont {Attaccalite},\ and\ \citenamefont {Olevano}}]{Bla11a}%
  \BibitemOpen
  \bibfield  {author} {\bibinfo {author} {\bibfnamefont {X.}~\bibnamefont
  {Blase}}, \bibinfo {author} {\bibfnamefont {C.}~\bibnamefont {Attaccalite}},
  \ and\ \bibinfo {author} {\bibfnamefont {V.}~\bibnamefont {Olevano}},\ }\href
  {\doibase 10.1103/PhysRevB.83.115103} {\bibfield  {journal} {\bibinfo
  {journal} {Phys. Rev. B}\ }\textbf {\bibinfo {volume} {83}},\ \bibinfo
  {pages} {115103} (\bibinfo {year} {2011})}\BibitemShut {NoStop}%
\bibitem [{\citenamefont {Krishnan}\ \emph {et~al.}(1980)\citenamefont
  {Krishnan}, \citenamefont {Binkley}, \citenamefont {Seeger},\ and\
  \citenamefont {Pople}}]{Pop80}%
  \BibitemOpen
  \bibfield  {author} {\bibinfo {author} {\bibfnamefont {R.}~\bibnamefont
  {Krishnan}}, \bibinfo {author} {\bibfnamefont {J.~S.}\ \bibnamefont
  {Binkley}}, \bibinfo {author} {\bibfnamefont {R.}~\bibnamefont {Seeger}}, \
  and\ \bibinfo {author} {\bibfnamefont {J.~A.}\ \bibnamefont {Pople}},\ }\href
  {\doibase 10.1063/1.438955} {\bibfield  {journal} {\bibinfo  {journal} {The
  Journal of Chemical Physics}\ }\textbf {\bibinfo {volume} {72}},\ \bibinfo
  {pages} {650} (\bibinfo {year} {1980})}\BibitemShut {NoStop}%
\bibitem [{\citenamefont {Weigend}(2006)}]{Wei06}%
  \BibitemOpen
  \bibfield  {author} {\bibinfo {author} {\bibfnamefont {F.}~\bibnamefont
  {Weigend}},\ }\href {\doibase 10.1039/B515623H} {\bibfield  {journal}
  {\bibinfo  {journal} {Phys. Chem. Chem. Phys.}\ }\textbf {\bibinfo {volume}
  {8}},\ \bibinfo {pages} {1057} (\bibinfo {year} {2006})}\BibitemShut
  {NoStop}%
\bibitem [{\citenamefont {Valiev}\ \emph {et~al.}(2010)\citenamefont {Valiev},
  \citenamefont {Bylaska}, \citenamefont {Govind}, \citenamefont {Kowalski},
  \citenamefont {Straatsma}, \citenamefont {Dam}, \citenamefont {Wang},
  \citenamefont {Nieplocha}, \citenamefont {Apra}, \citenamefont {Windus},\
  and\ \citenamefont {de~Jong}}]{nwchem}%
  \BibitemOpen
  \bibfield  {author} {\bibinfo {author} {\bibfnamefont {M.}~\bibnamefont
  {Valiev}}, \bibinfo {author} {\bibfnamefont {E.}~\bibnamefont {Bylaska}},
  \bibinfo {author} {\bibfnamefont {N.}~\bibnamefont {Govind}}, \bibinfo
  {author} {\bibfnamefont {K.}~\bibnamefont {Kowalski}}, \bibinfo {author}
  {\bibfnamefont {T.}~\bibnamefont {Straatsma}}, \bibinfo {author}
  {\bibfnamefont {H.~V.}\ \bibnamefont {Dam}}, \bibinfo {author} {\bibfnamefont
  {D.}~\bibnamefont {Wang}}, \bibinfo {author} {\bibfnamefont {J.}~\bibnamefont
  {Nieplocha}}, \bibinfo {author} {\bibfnamefont {E.}~\bibnamefont {Apra}},
  \bibinfo {author} {\bibfnamefont {T.}~\bibnamefont {Windus}}, \ and\ \bibinfo
  {author} {\bibfnamefont {W.}~\bibnamefont {de~Jong}},\ }\href {\doibase
  10.1016/j.cpc.2010.04.018} {\bibfield  {journal} {\bibinfo  {journal}
  {Comput. Phys. Comm.}\ }\textbf {\bibinfo {volume} {181}},\ \bibinfo {pages}
  {1477 } (\bibinfo {year} {2010})}\BibitemShut {NoStop}%
\bibitem [{\citenamefont {Sini}\ \emph {et~al.}(2011)\citenamefont {Sini},
  \citenamefont {Sears},\ and\ \citenamefont {Brédas}}]{Sini11}%
  \BibitemOpen
  \bibfield  {author} {\bibinfo {author} {\bibfnamefont {G.}~\bibnamefont
  {Sini}}, \bibinfo {author} {\bibfnamefont {J.~S.}\ \bibnamefont {Sears}}, \
  and\ \bibinfo {author} {\bibfnamefont {J.-L.}\ \bibnamefont {Brédas}},\
  }\href {\doibase 10.1021/ct1005517} {\bibfield  {journal} {\bibinfo
  {journal} {Journal of Chemical Theory and Computation}\ }\textbf {\bibinfo
  {volume} {7}},\ \bibinfo {pages} {602} (\bibinfo {year} {2011})}\BibitemShut
  {NoStop}%
\bibitem [{\citenamefont {Atalla}\ \emph {et~al.}(2013)\citenamefont {Atalla},
  \citenamefont {Yoon}, \citenamefont {Caruso}, \citenamefont {Rinke},\ and\
  \citenamefont {Scheffler}}]{Atalla13}%
  \BibitemOpen
  \bibfield  {author} {\bibinfo {author} {\bibfnamefont {V.}~\bibnamefont
  {Atalla}}, \bibinfo {author} {\bibfnamefont {M.}~\bibnamefont {Yoon}},
  \bibinfo {author} {\bibfnamefont {F.}~\bibnamefont {Caruso}}, \bibinfo
  {author} {\bibfnamefont {P.}~\bibnamefont {Rinke}}, \ and\ \bibinfo {author}
  {\bibfnamefont {M.}~\bibnamefont {Scheffler}},\ }\href {\doibase
  10.1103/PhysRevB.88.165122} {\bibfield  {journal} {\bibinfo  {journal} {Phys.
  Rev. B}\ }\textbf {\bibinfo {volume} {88}},\ \bibinfo {pages} {165122}
  (\bibinfo {year} {2013})}\BibitemShut {NoStop}%
\bibitem [{\citenamefont {Perdew}\ \emph {et~al.}(1996)\citenamefont {Perdew},
  \citenamefont {Ernzerhof},\ and\ \citenamefont {Burke}}]{Per96}%
  \BibitemOpen
  \bibfield  {author} {\bibinfo {author} {\bibfnamefont {J.~P.}\ \bibnamefont
  {Perdew}}, \bibinfo {author} {\bibfnamefont {M.}~\bibnamefont {Ernzerhof}}, \
  and\ \bibinfo {author} {\bibfnamefont {K.}~\bibnamefont {Burke}},\ }\href
  {\doibase 10.1063/1.472933} {\bibfield  {journal} {\bibinfo  {journal} {J.
  Chem. Phys.}\ }\textbf {\bibinfo {volume} {105}},\ \bibinfo {pages} {9982}
  (\bibinfo {year} {1996})}\BibitemShut {NoStop}%
\bibitem [{\citenamefont {Tsiper}\ and\ \citenamefont {Soos}(2001)}]{Tsi01}%
  \BibitemOpen
  \bibfield  {author} {\bibinfo {author} {\bibfnamefont {E.~V.}\ \bibnamefont
  {Tsiper}}\ and\ \bibinfo {author} {\bibfnamefont {Z.~G.}\ \bibnamefont
  {Soos}},\ }\href {\doibase 10.1103/PhysRevB.64.195124} {\bibfield  {journal}
  {\bibinfo  {journal} {Phys. Rev. B}\ }\textbf {\bibinfo {volume} {64}},\
  \bibinfo {pages} {195124} (\bibinfo {year} {2001})}\BibitemShut {NoStop}%
\bibitem [{\citenamefont {D{'}Avino}\ \emph {et~al.}(2014)\citenamefont
  {D{'}Avino}, \citenamefont {Muccioli}, \citenamefont {Zannoni}, \citenamefont
  {Beljonne},\ and\ \citenamefont {Soos}}]{Dav14}%
  \BibitemOpen
  \bibfield  {author} {\bibinfo {author} {\bibfnamefont {G.}~\bibnamefont
  {D{'}Avino}}, \bibinfo {author} {\bibfnamefont {L.}~\bibnamefont {Muccioli}},
  \bibinfo {author} {\bibfnamefont {C.}~\bibnamefont {Zannoni}}, \bibinfo
  {author} {\bibfnamefont {D.}~\bibnamefont {Beljonne}}, \ and\ \bibinfo
  {author} {\bibfnamefont {Z.~G.}\ \bibnamefont {Soos}},\ }\href {\doibase
  10.1021/ct500618w} {\bibfield  {journal} {\bibinfo  {journal} {J. Chem.
  Theory Comput.}\ }\textbf {\bibinfo {volume} {10}},\ \bibinfo {pages} {4959}
  (\bibinfo {year} {2014})}\BibitemShut {NoStop}%
\bibitem [{\citenamefont {Soos}\ \emph {et~al.}(2001)\citenamefont {Soos},
  \citenamefont {Tsiper},\ and\ \citenamefont {Pascal}}]{Tsi01_cpl}%
  \BibitemOpen
  \bibfield  {author} {\bibinfo {author} {\bibfnamefont {Z.}~\bibnamefont
  {Soos}}, \bibinfo {author} {\bibfnamefont {E.~V.}\ \bibnamefont {Tsiper}}, \
  and\ \bibinfo {author} {\bibfnamefont {R.}~\bibnamefont {Pascal}},\ }\href
  {\doibase 10.1016/S0009-2614(01)00661-3} {\bibfield  {journal} {\bibinfo
  {journal} {Chem. Phys. Lett.}\ }\textbf {\bibinfo {volume} {342}},\ \bibinfo
  {pages} {652 } (\bibinfo {year} {2001})}\BibitemShut {NoStop}%
\bibitem [{\citenamefont {D’Avino}\ \emph
  {et~al.}(2016{\natexlab{b}})\citenamefont {D’Avino}, \citenamefont
  {Vanzo},\ and\ \citenamefont {Soos}}]{Dav16_jcp}%
  \BibitemOpen
  \bibfield  {author} {\bibinfo {author} {\bibfnamefont {G.}~\bibnamefont
  {D’Avino}}, \bibinfo {author} {\bibfnamefont {D.}~\bibnamefont {Vanzo}}, \
  and\ \bibinfo {author} {\bibfnamefont {Z.~G.}\ \bibnamefont {Soos}},\ }\href
  {\doibase 10.1063/1.4939840} {\bibfield  {journal} {\bibinfo  {journal} {The
  Journal of Chemical Physics}\ }\textbf {\bibinfo {volume} {144}},\ \bibinfo
  {pages} {034702} (\bibinfo {year} {2016}{\natexlab{b}})}\BibitemShut
  {NoStop}%
\bibitem [{\citenamefont {Zerner}\ \emph {et~al.}(1980)\citenamefont {Zerner},
  \citenamefont {Loew}, \citenamefont {Kirchner},\ and\ \citenamefont
  {Mueller-Westerhoff}}]{zindo}%
  \BibitemOpen
  \bibfield  {author} {\bibinfo {author} {\bibfnamefont {M.~C.}\ \bibnamefont
  {Zerner}}, \bibinfo {author} {\bibfnamefont {G.~H.}\ \bibnamefont {Loew}},
  \bibinfo {author} {\bibfnamefont {R.~F.}\ \bibnamefont {Kirchner}}, \ and\
  \bibinfo {author} {\bibfnamefont {U.~T.}\ \bibnamefont
  {Mueller-Westerhoff}},\ }\href {\doibase 10.1021/ja00522a025} {\bibfield
  {journal} {\bibinfo  {journal} {J. Am. Chem. Soc.}\ }\textbf {\bibinfo
  {volume} {102}},\ \bibinfo {pages} {589} (\bibinfo {year}
  {1980})}\BibitemShut {NoStop}%
\bibitem [{\citenamefont {Terenziani}\ \emph {et~al.}(2006)\citenamefont
  {Terenziani}, \citenamefont {Painelli}, \citenamefont {Katan}, \citenamefont
  {Charlot},\ and\ \citenamefont {Blanchard-Desce}}]{terenziani06}%
  \BibitemOpen
  \bibfield  {author} {\bibinfo {author} {\bibfnamefont {F.}~\bibnamefont
  {Terenziani}}, \bibinfo {author} {\bibfnamefont {A.}~\bibnamefont
  {Painelli}}, \bibinfo {author} {\bibfnamefont {C.}~\bibnamefont {Katan}},
  \bibinfo {author} {\bibfnamefont {M.}~\bibnamefont {Charlot}}, \ and\
  \bibinfo {author} {\bibfnamefont {M.}~\bibnamefont {Blanchard-Desce}},\
  }\href {\doibase 10.1021/ja064521j} {\bibfield  {journal} {\bibinfo
  {journal} {J. Am. Chem. Soc.}\ }\textbf {\bibinfo {volume} {128}},\ \bibinfo
  {pages} {15742} (\bibinfo {year} {2006})}\BibitemShut {NoStop}%
\bibitem [{\citenamefont {Terenziani}\ \emph {et~al.}(2008)\citenamefont
  {Terenziani}, \citenamefont {Sissa},\ and\ \citenamefont
  {Painelli}}]{terenziani08}%
  \BibitemOpen
  \bibfield  {author} {\bibinfo {author} {\bibfnamefont {F.}~\bibnamefont
  {Terenziani}}, \bibinfo {author} {\bibfnamefont {C.}~\bibnamefont {Sissa}}, \
  and\ \bibinfo {author} {\bibfnamefont {A.}~\bibnamefont {Painelli}},\ }\href
  {\doibase 10.1021/jp710241g} {\bibfield  {journal} {\bibinfo  {journal} {J.
  Phys. Chem. B}\ }\textbf {\bibinfo {volume} {112}},\ \bibinfo {pages} {5079}
  (\bibinfo {year} {2008})}\BibitemShut {NoStop}%
\bibitem [{\citenamefont {D’Avino}\ \emph
  {et~al.}(2016{\natexlab{c}})\citenamefont {D’Avino}, \citenamefont
  {Muccioli}, \citenamefont {Olivier},\ and\ \citenamefont
  {Beljonne}}]{Dav16opv}%
  \BibitemOpen
  \bibfield  {author} {\bibinfo {author} {\bibfnamefont {G.}~\bibnamefont
  {D’Avino}}, \bibinfo {author} {\bibfnamefont {L.}~\bibnamefont {Muccioli}},
  \bibinfo {author} {\bibfnamefont {Y.}~\bibnamefont {Olivier}}, \ and\
  \bibinfo {author} {\bibfnamefont {D.}~\bibnamefont {Beljonne}},\ }\href
  {\doibase 10.1021/acs.jpclett.5b02680} {\bibfield  {journal} {\bibinfo
  {journal} {J. Phys. Chem. Lett.}\ }\textbf {\bibinfo {volume} {7}},\ \bibinfo
  {pages} {536} (\bibinfo {year} {2016}{\natexlab{c}})}\BibitemShut {NoStop}%
\bibitem [{\citenamefont {Valeev}\ \emph {et~al.}(2006)\citenamefont {Valeev},
  \citenamefont {Coropceanu}, \citenamefont {da~Silva~Filho}, \citenamefont
  {Salman},\ and\ \citenamefont {Brédas}}]{valeev06}%
  \BibitemOpen
  \bibfield  {author} {\bibinfo {author} {\bibfnamefont {E.~F.}\ \bibnamefont
  {Valeev}}, \bibinfo {author} {\bibfnamefont {V.}~\bibnamefont {Coropceanu}},
  \bibinfo {author} {\bibfnamefont {D.~A.}\ \bibnamefont {da~Silva~Filho}},
  \bibinfo {author} {\bibfnamefont {S.}~\bibnamefont {Salman}}, \ and\ \bibinfo
  {author} {\bibfnamefont {J.-L.}\ \bibnamefont {Brédas}},\ }\href {\doibase
  10.1021/ja061827h} {\bibfield  {journal} {\bibinfo  {journal} {J. Am. Chem.
  Soc.}\ }\textbf {\bibinfo {volume} {128}},\ \bibinfo {pages} {9882} (\bibinfo
  {year} {2006})}\BibitemShut {NoStop}%
\bibitem [{\citenamefont {Cave}\ and\ \citenamefont {Newton}(1996)}]{cave96}%
  \BibitemOpen
  \bibfield  {author} {\bibinfo {author} {\bibfnamefont {R.~J.}\ \bibnamefont
  {Cave}}\ and\ \bibinfo {author} {\bibfnamefont {M.~D.}\ \bibnamefont
  {Newton}},\ }\href {\doibase 10.1016/0009-2614(95)01310-5} {\bibfield
  {journal} {\bibinfo  {journal} {Chem. Phys. Lett.}\ }\textbf {\bibinfo
  {volume} {249}},\ \bibinfo {pages} {15} (\bibinfo {year} {1996})}\BibitemShut
  {NoStop}%
\bibitem [{\citenamefont {Hellweg}\ \emph {et~al.}(2008)\citenamefont
  {Hellweg}, \citenamefont {Grun},\ and\ \citenamefont {Hattig}}]{hellweg08}%
  \BibitemOpen
  \bibfield  {author} {\bibinfo {author} {\bibfnamefont {A.}~\bibnamefont
  {Hellweg}}, \bibinfo {author} {\bibfnamefont {S.~A.}\ \bibnamefont {Grun}}, \
  and\ \bibinfo {author} {\bibfnamefont {C.}~\bibnamefont {Hattig}},\ }\href
  {\doibase 10.1039/B803727B} {\bibfield  {journal} {\bibinfo  {journal} {Phys.
  Chem. Chem. Phys.}\ }\textbf {\bibinfo {volume} {10}},\ \bibinfo {pages}
  {4119} (\bibinfo {year} {2008})}\BibitemShut {NoStop}%
\bibitem [{\citenamefont {Siegrist}\ \emph {et~al.}(2001)\citenamefont
  {Siegrist}, \citenamefont {Kloc}, \citenamefont {Sch\"{o}n}, \citenamefont
  {Batlogg}, \citenamefont {Haddon}, \citenamefont {Berg},\ and\ \citenamefont
  {Thomas}}]{Sie01}%
  \BibitemOpen
  \bibfield  {author} {\bibinfo {author} {\bibfnamefont {T.}~\bibnamefont
  {Siegrist}}, \bibinfo {author} {\bibfnamefont {C.}~\bibnamefont {Kloc}},
  \bibinfo {author} {\bibfnamefont {J.~H.}\ \bibnamefont {Sch\"{o}n}}, \bibinfo
  {author} {\bibfnamefont {B.}~\bibnamefont {Batlogg}}, \bibinfo {author}
  {\bibfnamefont {R.~C.}\ \bibnamefont {Haddon}}, \bibinfo {author}
  {\bibfnamefont {S.}~\bibnamefont {Berg}}, \ and\ \bibinfo {author}
  {\bibfnamefont {G.~A.}\ \bibnamefont {Thomas}},\ }\href {\doibase
  10.1002/1521-3773(20010504)40:9<1732::AID-ANIE17320>3.0.CO;2-7} {\bibfield
  {journal} {\bibinfo  {journal} {Angew. Chem. Int. Ed.}\ }\textbf {\bibinfo
  {volume} {40}},\ \bibinfo {pages} {1732} (\bibinfo {year}
  {2001})}\BibitemShut {NoStop}%
\bibitem [{\citenamefont {Salzmann}()}]{ingo}%
  \BibitemOpen
  \bibfield  {author} {\bibinfo {author} {\bibfnamefont {I.}~\bibnamefont
  {Salzmann}},\ }\href@noop {} {}\bibinfo {howpublished} {Private
  communication}\BibitemShut {NoStop}%
\bibitem [{\citenamefont {Sakanoue}\ and\ \citenamefont
  {Sirringhaus}(2010)}]{Sak10}%
  \BibitemOpen
  \bibfield  {author} {\bibinfo {author} {\bibfnamefont {T.}~\bibnamefont
  {Sakanoue}}\ and\ \bibinfo {author} {\bibfnamefont {H.}~\bibnamefont
  {Sirringhaus}},\ }\href {\doibase 10.1038/nmat2825} {\bibfield  {journal}
  {\bibinfo  {journal} {Nat. Mater.}\ }\textbf {\bibinfo {volume} {9}},\
  \bibinfo {pages} {736} (\bibinfo {year} {2010})}\BibitemShut {NoStop}%
\bibitem [{\citenamefont {Winkler}\ \emph {et~al.}(2015)\citenamefont
  {Winkler}, \citenamefont {Amsalem}, \citenamefont {Frisch}, \citenamefont
  {Oehzelt}, \citenamefont {Heimel},\ and\ \citenamefont {Koch}}]{Win15}%
  \BibitemOpen
  \bibfield  {author} {\bibinfo {author} {\bibfnamefont {S.}~\bibnamefont
  {Winkler}}, \bibinfo {author} {\bibfnamefont {P.}~\bibnamefont {Amsalem}},
  \bibinfo {author} {\bibfnamefont {J.}~\bibnamefont {Frisch}}, \bibinfo
  {author} {\bibfnamefont {M.}~\bibnamefont {Oehzelt}}, \bibinfo {author}
  {\bibfnamefont {G.}~\bibnamefont {Heimel}}, \ and\ \bibinfo {author}
  {\bibfnamefont {N.}~\bibnamefont {Koch}},\ }\href {\doibase
  10.1039/C5MH00023H} {\bibfield  {journal} {\bibinfo  {journal} {Mater.
  Horiz.}\ }\textbf {\bibinfo {volume} {2}},\ \bibinfo {pages} {427} (\bibinfo
  {year} {2015})}\BibitemShut {NoStop}%
\bibitem [{\citenamefont {Tsutsui}\ \emph {et~al.}(2016)\citenamefont
  {Tsutsui}, \citenamefont {Schweicher}, \citenamefont {Chattopadhyay},
  \citenamefont {Sakurai}, \citenamefont {Arlin}, \citenamefont {Ruzié},
  \citenamefont {Aliev}, \citenamefont {Ciesielski}, \citenamefont {Colella},
  \citenamefont {Kennedy}, \citenamefont {Lemaur}, \citenamefont {Olivier},
  \citenamefont {Hadji}, \citenamefont {Sanguinet}, \citenamefont {Castet},
  \citenamefont {Osella}, \citenamefont {Dudenko}, \citenamefont {Beljonne},
  \citenamefont {Cornil}, \citenamefont {Samorì}, \citenamefont {Seki},\ and\
  \citenamefont {Geerts}}]{Tsu16}%
  \BibitemOpen
  \bibfield  {author} {\bibinfo {author} {\bibfnamefont {Y.}~\bibnamefont
  {Tsutsui}}, \bibinfo {author} {\bibfnamefont {G.}~\bibnamefont {Schweicher}},
  \bibinfo {author} {\bibfnamefont {B.}~\bibnamefont {Chattopadhyay}}, \bibinfo
  {author} {\bibfnamefont {T.}~\bibnamefont {Sakurai}}, \bibinfo {author}
  {\bibfnamefont {J.-B.}\ \bibnamefont {Arlin}}, \bibinfo {author}
  {\bibfnamefont {C.}~\bibnamefont {Ruzié}}, \bibinfo {author} {\bibfnamefont
  {A.}~\bibnamefont {Aliev}}, \bibinfo {author} {\bibfnamefont
  {A.}~\bibnamefont {Ciesielski}}, \bibinfo {author} {\bibfnamefont
  {S.}~\bibnamefont {Colella}}, \bibinfo {author} {\bibfnamefont {A.~R.}\
  \bibnamefont {Kennedy}}, \bibinfo {author} {\bibfnamefont {V.}~\bibnamefont
  {Lemaur}}, \bibinfo {author} {\bibfnamefont {Y.}~\bibnamefont {Olivier}},
  \bibinfo {author} {\bibfnamefont {R.}~\bibnamefont {Hadji}}, \bibinfo
  {author} {\bibfnamefont {L.}~\bibnamefont {Sanguinet}}, \bibinfo {author}
  {\bibfnamefont {F.}~\bibnamefont {Castet}}, \bibinfo {author} {\bibfnamefont
  {S.}~\bibnamefont {Osella}}, \bibinfo {author} {\bibfnamefont
  {D.}~\bibnamefont {Dudenko}}, \bibinfo {author} {\bibfnamefont
  {D.}~\bibnamefont {Beljonne}}, \bibinfo {author} {\bibfnamefont
  {J.}~\bibnamefont {Cornil}}, \bibinfo {author} {\bibfnamefont
  {P.}~\bibnamefont {Samorì}}, \bibinfo {author} {\bibfnamefont
  {S.}~\bibnamefont {Seki}}, \ and\ \bibinfo {author} {\bibfnamefont {Y.~H.}\
  \bibnamefont {Geerts}},\ }\href {\doibase 10.1002/adma.201601285} {\bibfield
  {journal} {\bibinfo  {journal} {Adv. Mater.}\ }\textbf {\bibinfo {volume}
  {28}},\ \bibinfo {pages} {7106} (\bibinfo {year} {2016})}\BibitemShut
  {NoStop}%
\bibitem [{\citenamefont {Delchiaro}\ \emph {et~al.}(2017)\citenamefont
  {Delchiaro}, \citenamefont {Girlando}, \citenamefont {Painelli},
  \citenamefont {Bandyopadhyay}, \citenamefont {Pati},\ and\ \citenamefont
  {D'Avino}}]{Delc17}%
  \BibitemOpen
  \bibfield  {author} {\bibinfo {author} {\bibfnamefont {F.}~\bibnamefont
  {Delchiaro}}, \bibinfo {author} {\bibfnamefont {A.}~\bibnamefont {Girlando}},
  \bibinfo {author} {\bibfnamefont {A.}~\bibnamefont {Painelli}}, \bibinfo
  {author} {\bibfnamefont {A.}~\bibnamefont {Bandyopadhyay}}, \bibinfo {author}
  {\bibfnamefont {S.~K.}\ \bibnamefont {Pati}}, \ and\ \bibinfo {author}
  {\bibfnamefont {G.}~\bibnamefont {D'Avino}},\ }\href {\doibase
  10.1103/PhysRevB.95.155125} {\bibfield  {journal} {\bibinfo  {journal} {Phys.
  Rev. B}\ }\textbf {\bibinfo {volume} {95}},\ \bibinfo {pages} {155125}
  (\bibinfo {year} {2017})}\BibitemShut {NoStop}%
\bibitem [{\citenamefont {Colle}\ \emph {et~al.}(2011)\citenamefont {Colle},
  \citenamefont {Grosso}, \citenamefont {Ronzani},\ and\ \citenamefont
  {Zicovich-Wilson}}]{Col11}%
  \BibitemOpen
  \bibfield  {author} {\bibinfo {author} {\bibfnamefont {R.}~\bibnamefont
  {Colle}}, \bibinfo {author} {\bibfnamefont {G.}~\bibnamefont {Grosso}},
  \bibinfo {author} {\bibfnamefont {A.}~\bibnamefont {Ronzani}}, \ and\
  \bibinfo {author} {\bibfnamefont {C.~M.}\ \bibnamefont {Zicovich-Wilson}},\
  }\href {\doibase 10.1002/pssb.201046429} {\bibfield  {journal} {\bibinfo
  {journal} {Phys. Status Solidi b}\ }\textbf {\bibinfo {volume} {248}},\
  \bibinfo {pages} {1360} (\bibinfo {year} {2011})}\BibitemShut {NoStop}%
\end{thebibliography}%

\end{document}